\documentstyle[12pt]{article}

\def\hybrid{\topmargin -20pt	\oddsidemargin 0pt
	\headheight 0pt	\headsep 0pt
        \textwidth 6.35in
        \textheight 9.65in
	\marginparwidth .875in
	\parskip 5pt plus 1pt	\jot = 1.5ex}

\def\baselinestretch{1.2}

\catcode`\@=11

\def\marginnote#1{}
\newcount\hour
\newcount\minute
\newtoks\amorpm
\hour=\time\divide\hour by60
\minute=\time{\multiply\hour by60 \global\advance\minute by-\hour}
\edef\standardtime{{\ifnum\hour<12 \global\amorpm={am}%
	\else\global\amorpm={pm}\advance\hour by-12 \fi
	\ifnum\hour=0 \hour=12 \fi
	\number\hour:\ifnum\minute<10 0\fi\number\minute\the\amorpm}}
\edef\militarytime{\number\hour:\ifnum\minute<10 0\fi\number\minute}
\def\draftlabel#1{{\@bsphack\if@filesw {\let\thepage\relax
   \xdef\@gtempa{\write\@auxout{\string
      \newlabel{#1}{{\@currentlabel}{\thepage}}}}}\@gtempa
   \if@nobreak \ifvmode\nobreak\fi\fi\fi\@esphack}
	\gdef\@eqnlabel{#1}}
\def\@eqnlabel{}
\def\@vacuum{}
\def\draftmarginnote#1{\marginpar{\raggedright\scriptsize\tt#1}}

\def\draft{\oddsidemargin -.2truein
	\def\@oddfoot{\sl preliminary draft \hfil
	\rm\thepage\hfil\sl\today\quad\militarytime}
	\let\@evenfoot\@oddfoot	\overfullrule 3pt
	\let\label=\draftlabel
	\let\marginnote=\draftmarginnote
   \def\@eqnnum{(\theequation)\rlap{\kern\marginparsep\tt\@eqnlabel}%
\global\let\@eqnlabel\@vacuum}  }

\def\preprint{\twocolumn\sloppy\flushbottom\parindent 2em
	\leftmargini 2em\leftmarginv .5em\leftmarginvi .5em
	\oddsidemargin -.5in	\evensidemargin -.5in
	\columnsep .4in	\footheight 0pt
	\textwidth 10.in	\topmargin  -.4in
	\headheight 12pt \topskip .4in
	\textheight 6.9in \footskip 0pt
	\def\@oddhead{\thepage\hfil\addtocounter{page}{1}\thepage}
	\let\@evenhead\@oddhead	\def\@oddfoot{}	\def\@evenfoot{} }

\def\numberbysection{\@addtoreset{equation}{section}
	\def\theequation{\thesection.\arabic{equation}}}

\def\underline#1{\relax\ifmmode\@@underline#1\else
	$\@@underline{\hbox{#1}}$\relax\fi}

\def\titlepage{\@restonecolfalse\if@twocolumn\@restonecoltrue
\onecolumn
     \else \newpage \fi \thispagestyle{empty}\c@page\z@
	\def\thefootnote{\fnsymbol{footnote}} }

\def\endtitlepage{\if@restonecol\twocolumn \else \newpage \fi
	\def\thefootnote{\arabic{footnote}}
	\setcounter{footnote}{0}}  

\catcode`@=12
\relax

\def\figcap{\section*{Figure Captions\markboth
	{FIGURECAPTIONS}{FIGURECAPTIONS}}\list
	{Figure \arabic{enumi}:\hfill}{\settowidth\labelwidth{Figure  
999:}
	\leftmargin\labelwidth
	\advance\leftmargin\labelsep\usecounter{enumi}}}
 \relax
\def\tablecap{\section*{Table Captions\markboth
	{TABLECAPTIONS}{TABLECAPTIONS}}\list
	{Table \arabic{enumi}:\hfill}{\settowidth\labelwidth{Table  
999:}
	\leftmargin\labelwidth
	\advance\leftmargin\labelsep\usecounter{enumi}}}
 \relax
\def\reflist{\section*{References\markboth
	{REFLIST}{REFLIST}}\list
	{[\arabic{enumi}]\hfill}{\settowidth\labelwidth{[999]}
	\leftmargin\labelwidth
	\advance\leftmargin\labelsep\usecounter{enumi}}}
 \relax
\makeatletter
\newcounter{pubctr}
\def\publist{\@ifnextchar[{\@publist}{\@@publist}}
\def\@publist[#1]{\list
	{[\arabic{pubctr}]\hfill}{\settowidth\labelwidth{[999]}
	\leftmargin\labelwidth
	\advance\leftmargin\labelsep
	\@nmbrlisttrue\def\@listctr{pubctr}
	\setcounter{pubctr}{#1}\addtocounter{pubctr}{-1}}}
\def\@@publist{\list
	{[\arabic{pubctr}]\hfill}{\settowidth\labelwidth{[999]}
	\leftmargin\labelwidth
	\advance\leftmargin\labelsep
	\@nmbrlisttrue\def\@listctr{pubctr}}}
 \relax
\makeatother
\newskip\humongous \humongous=0pt plus 1000pt minus 1000pt

\newif\ifdtup

\relax
\hybrid
\def\lrp{\stackrel{\leftrightarrow}{\partial}}

\def\thefootnote{\fnsymbol{footnote}}
\def\be{\begin{equation}}
\def\ee{\end{equation}}
\def\ba{\begin{eqnarray}}
\def\ea{\end{eqnarray}}
\def\a{\alpha}
\def\b{\beta}

\def\th{\vartheta}

\def\e{\varepsilon}
\def\l{\lambda}

\def\im{{\rm Im}\tau}

\font\fivesans=cmss10 at 4.61pt
\font\sevensans=cmss10 at 6.81pt
\font\tensans=cmss10
\newfam\sansfam
\textfont\sansfam=\tensans\scriptfont\sansfam=
\sevensans\scriptscriptfont
\sansfam=\fivesans
\def\sans{\fam\sansfam\tensans}

\def\Z{{\mathchoice
{\hbox{$\sans\textstyle Z\kern-0.4em Z$}}
{\hbox{$\sans\textstyle Z\kern-0.4em Z$}}
{\hbox{$\sans\scriptstyle Z\kern-0.3em Z$}}
{\hbox{$\sans\scriptscriptstyle Z\kern-0.2em Z$}}}}
\newlength{\boxsize} 

\def\lrp{\stackrel{\leftrightarrow}{\partial}}

\begin{document}
\renewcommand{\theequation}{\thesection.\arabic{equation}}
\newcommand{\beq}{\begin{equation}}
\newcommand{\eeq}[1]{\label{#1}\end{equation}}
\newcommand{\ber}{\begin{eqnarray}}
\newcommand{\eer}[1]{\label{#1}\end{eqnarray}}
\begin{titlepage}
\begin{center}

\hfill CERN-TH/97-39\\
\hfill LPTENS-97/10\\
\hfill hep-th/9703059\\
\vskip .2in

{\large \bf  Perturbative and Non-Perturbative Partial Supersymmetry
Breaking:
 $N=4\rightarrow N=2\rightarrow N=1$}
\vskip .5in

{\bf Elias Kiritsis and Costas Kounnas\footnote{On leave from Ecole
Normale Sup\'erieure, 24 rue Lhomond, F-75231, Paris Cedex 05,
FRANCE.}}\\
\vskip
 .1in
{\em Theory Division, CERN,\\ CH-1211,
Geneva 23, SWITZERLAND} \footnote{e-mail addresses:
KIRITSIS,KOUNNAS@NXTH04.CERN.CH}\\

\end{center}

\begin{center} {\bf ABSTRACT }
 \end{center}
\begin{quotation}\noindent
\baselineskip 10pt

We show the existence of a supersymmetry breaking mechanism in
string theory, where $N=4$ supersymmetry is broken
{\em spontaneously} to $N=2$  and $N=1$ with moduli dependent gravitino
masses. The spectrum of the spontaneously broken theory with lower
supersymmetry is in one-to-one correspondence
with the spectrum of the heterotic $N=4$ string. The mass splitting of
the $N=4$ spectrum depends on the compactification moduli as well as
the three $R$-symmetry charges.
We also show that, in string theory, chiral theories can be obtained
after spontaneous breaking of extended supersymmetry.
This was impossible at the level of field theory.

In the large  moduli limit a restoration of the $N=4$
supersymmetry is obtained.   As expected the graviphotons and some of
the gauge
bosons become massive in $N=1$ vacua. At some special points of the
moduli space some of the $N=4$ states with  non-zero winding numbers
and  with spin 0 and {1/2}  become massless chiral superfields of the
unbroken $N=1$ supersymmetry. Such vacua have a dual type II
description, in which
there are magnetically charged states with spin 0 and {1/2} that become
massless.
The heterotic-type II duality suggests some novel non-perturbative
transitions
on the type II side. Such transitions do not seem to have a geometric
interpretation, since they relate type II vacua with symmetric
worldsheet
structure to asymmetric ones.
The heterotic interpretation of such a transition is an ordinary
higgsing of an SU(2) factor.

In the case of $N=4 \rightarrow N=2$, the perturbative $N=2$
prepotential is determined by the perturbative $N=4$ BPS states. This
observation permits us to  suggest a  method to determine the exact
non-perturbative prepotential of the effective $N=2$ supergravity using
the shifted spectrum  of the  non-perturbative BPS states of the
underlying $N=4$ theory.

\end{quotation}
\vskip .2in
CERN-TH/97-39\\
March 1997\\
\end{titlepage}
\vfill
\eject
\def\baselinestretch{1.2}
\baselineskip 14 pt

\noindent
\section{Introduction}
\setcounter{equation}{0}

When a local symmetry is spontaneously broken, the physical states
can
be classified in terms of the unbroken phase spectrum and in terms of
a well-defined mass splitting given in terms of vacuum expectation
values of some fields,
weighted by the charges of the broken symmetry. In the case of
gauge symmetry breaking, the fields with non-zero vev's are physical
scalar fields, while in the case of supersymmetry breaking they are
auxiliary fields. In extended supersymmetric theories (local or
global), the supersymmetric vacua are  degenerate, with zero vacuum
energy for any vev of the moduli fields ($S,T^i$). For instance, in
the case of $N=4$ supergravity based on a gauge group
$U(1)^6\times~G$, the space of the moduli fields is given in terms of
    $2+6r$ physical scalars, which are coordinates  of the coset
space \cite{{Roo},{fgkp}}
\be
\left[{SU(1,1)\over U(1)}\right]_S\times \left[{SO(6, r)\over
{SO(6)\times SO(r)}}\right]_{T}
\label{moduli}
\ee
$r$ is the rank of the gauge group $G$.

In an arbitrary point of the moduli space the gauge symmetry $G$ is
broken down to $U(1)^r$ while at some special points of the moduli
space the gauge symmetry
is extended to some non-Abelian gauge group of  the same rank due to
the presence of some extra gauge multiplets that become massless at
the  special points of the moduli space.

In the heterotic $N=4$ superstring solution obtained by $T^6$
compactification of the  ten dimensional superstring, the rank of the
group $r$ has a fixed value, $r=22$ \cite{nsw}--\cite{lls}. In an
arbitrary point of the
moduli space the gauge group is $U(1)^r$
and in special points the symmetry is extended as in field theory.
There is however a fundamental difference between the field theory
Higgs phenomenon and the string theory  one. Indeed, if in  an $N=4$
field theory the gauge group is $G=U(1)^6\times SO(32)$ at any given
point of the
moduli space, then at any other point the remaining gauge symmetry
$G_{T_i} $ is always a subgroup of  $G$ with smaller dimensionality
dim$(G_{T_i})\le$ dim($G$). On the contrary, in the string Higgs
phenomenon, owing to the existence of winding states, we can connect
gauge groups
which are not subgroups of a larger  gauge group. For instance, it is
possible to connect continuously $G= U(1)^6\times SO(32)$ with  $G=
U(1)^6\times E_8\times E_8$, as well as with the most symmetric one of
the same rank, namely  $G=SO(44)$.
Indeed, starting from a ten dimensional $N=1$ supergravity theory with
$G=SO(32)$ or $G=E_8\times E_8$ after compactification in four
dimensions the only possible $N=4$ supergravity effective theories
are
based either to $G=U(1)^6\times SO(32)$ or $G=U(1)^6\times E_8\times
E_8$ (and their subgroups obtained with Higgs phenomenon). In string
theory the gauge group can be further extended due to the existence
of extra gauge bosons with non-zero winding numbers, which can become
massless in special points of the moduli space.

When some auxiliary fields of the supergravity theories have
a non-vanishing
vev, some (or all) of the supersymmetries are spontaneously broken
\cite{ss1}--\cite{kr}.
There is a consistent class of $N=1,2$ and $N=4$ models defined in
flat space-time in which all supersymmetries are broken or partially
broken \cite{abk}--\cite{kr}. The most interesting case for our
purposes is that in which there is one of the
supersymmetries left unbroken. In that case we know that it is
possible, in general, to have chiral representations of matter scalar
multiplets
which can describe the quarks and leptons of the supersymmetric
standard model.  All previous examples about the
partial breaking of $N=2$ to $N=1$ supersymmetry was done at the
field theory level \cite{apt}. In this work we will first show the
extension of the partial spontaneous breaking at the perturbative
string level and then we will generalize our result to the
non-perturbative level using as a tool the heterotic--type II string
duality of the N=4 four dimensional superstrings
\cite{{duff},{HT},{wi}}.

In the process we will present evidence that non-perturbatively string
ground states are of the spontaneously broken kind, the
massive gravitini sometimes being solitons.
We will also be able to find evidence for some novel non-perturbative
transition in string theory between symmetric (geometric) and
asymmetric
(non-geometric) compactifications.

Moreover we construct examples of string ground states with
spontaneously broken
N=2$\to$N=1 supersymmetry and chiral N=1 spectrum.
This shows that, unlike field theory, chirality can appear during
spontaneous breaking of extended supersymmetry in string theory.

The structure of the present paper is as follows:
In section 2 we review the perturbative N=4 spectrum of string
theories.
Based on N=4 supergravity, we describe the (non-perturbative) BPS mass
formula
and analyse its moduli dependence.

In section 3 we describe heterotic ground states with $N=4\to N=2$
spontaneously broken supersymmetry.
We give two complementary descriptions: one in terms of specific
freely acting orbifolds and another in terms of generalized Lorentzian
boosts of the unbroken theory at a special value in moduli space.
We analyse and compare the behaviour of thresholds in such ground
states
to those of conventional $K_3\times T^2$ compactifications.

In section 4 we discuss the spectrum of perturbative BPS states for
heterotic ground-states with $N=4\to N=2$ spontaneously broken
supersymmetry.
We calculate the BPS multiplicities and by analysing  the BPS mass
formula
we describe the points in moduli space where BPS states become
massless.

In section 5 we describe (partial) spontaneous $N=4\to N=1$
supersymmetry breaking. We construct as an example a heterotic model
that realizes this breaking pattern, and calculate its thresholds.
Similarly in section 6 we describe spontaneous $N=2\to N=1$
supersymmetry breaking and provide as an example a heterotic ground
state that realizes
this possibility with chiral massless spectrum.
Its gauge thresholds are also computed.

In section 7 we construct and analyse the type II duals of the
heterotic models with spontaneously broken  $N=4\to N=2$ spacetime
supersymmetry. Evidence is presented for non-geometric non-perturbative
transitions in the type II side which correspond to the ordinary Higgs
effect
on the heterotic side.

Finally in section 8 we present a conjecture on the non-perturbative
BPS spectrum on the string ground states with $N=4\to N=2$ described in
this paper.

\section{Perturbative and Non-Perturbative $N=4$ Mass Spectrum}
\setcounter{equation}{0}

Our starting point is a four dimensional heterotic $N=4$ superstring
solution.
{}From  the world-sheet viewpoint these theories are constructed by
the following left- and right- moving degrees of freedom:

$\bullet$ Four left-moving non-compact super-coordinates,
{\bf $X^{\mu}, \Psi^{\mu}$}

$\bullet$ Six left-moving compactified supercoordinates,
{\bf $\Phi^{I}, \Psi^{I}$}

$\bullet$ The left-moving super-reparametrization ghosts,
{\bf $b, c$} and {\bf $\beta, \gamma$}

$\bullet$ Four right-moving  coordinates,
{\bf  ${\bar X}^{\mu}$}

$\bullet$ Six right-moving compactified coordinates,
{\bf  ${\bar \Phi}^{I}$}

$\bullet$ 32 right-moving fermions,
{\bf ${\bar \Psi}^A$}

$\bullet$ The right-moving reparametrization ghosts,
{\bf $\bar b, \bar c$}

In order to obtain consistent (without ghosts) $N=4$ solution the
left-moving fermions
$\Psi^{\mu}, \Psi^{I}$ and the $\beta, \gamma$ ghosts must have the
same boundary conditions. In that case the global existence of the
left-moving spin-$3/2$ world-sheet supercurrent
\be
T_F~=~\Psi^{\mu}\partial X^{\mu} + \Psi^{I}\partial\Phi^I
\label{supcur}
\ee
implies periodic boundary conditions for the compact and non-compact
left-moving coordinates, $\Phi^I,~ X^{\mu}$. Modular invariance
implies the right-moving coordinates ${\bar \Phi}^I,~ {\bar
X}^{\mu}$ to be periodic as well.
The solution with $G=U(1)^6\times SO(32)$ is when the right- moving
fermions $\Psi^{A}$ have the same boundary conditions (periodic or
antiperiodic), while the solution with $G=U(1)^6\times E_8\times E_8$
is when the ${\bar\Psi}^{A}=({\bar \Psi}^{A_1},~ {\bar \Psi}^{A_2})$
are in two groups of sixteen with the same boundary conditions.
Starting either from the $G=U(1)^6\times E_8\times E_8$ solution  or
from the $G=U(1)^6\times SO(32)$  we can obtain all others by
deforming the momentum lattice of compactified bosons together with
the charge lattice of the 32 fermions ${\bar\Psi}^{A}$.

The  partition function of the heterotic $N=4$ solutions in a generic
point of the moduli space is well known and has the following
expression:
\be
Z(\tau ,\bar\tau)={1\over \tau_2|\eta|^4}~{1\over 2}
\sum_{\alpha,\beta}
(-)^{\alpha+\beta +\alpha \beta}~
\frac{\th^{4}[^{\alpha}_{\beta}](\tau)}{\eta^4(\tau)}~
 {\Gamma_{(6,22)}(\tau,{\bar\tau})\over \eta^6\bar\eta^{22}}
\label{n4}
\ee
where $ \Gamma_{(6,22)}(\tau,{\bar\tau})$ denotes the partition
function
due to the compactified coordinates $\Phi^I, {\bar \Phi}^A$ and due
to the sixteen right-moving $U(1)$ currents constructed with the
fermions ${\bar \Psi^I}$
\be
{\bar J}^k= {\bar \Psi}^{2k-1}{\bar \Psi}^{2k},~~~~k=1,2,\cdots,16.
\label{b1}\ee
$\Gamma_{(6,22)}(\tau,{\bar\tau})$ has in total $6\times 22$ moduli
parameters which correspond to (1,1) marginal deformations
of the world-sheet action:
\be
\delta S^{2d}~=~\delta T_{IJ}~\partial \Phi^I~\partial {\bar \Phi}^J
{}~+~
Y_{I}^{k}~\partial \Phi^I~{\bar J}^k
\label{b2}\ee

In terms of the six dimensional backgrounds of the compactified
space, the $T_{IJ}$ moduli are related to the internal background
metric
$G_{IJ}$ and  the  internal antisymmetric tensor $B_{IJ}$; $T_{IJ}$
=$G_{IJ}$+ $B_{IJ}$. The $Y_{I}^{k}$ moduli are the six
dimensional
internal gauge fields backgrounds which belong in the Cartan
subalgebra of the ten-dimensional gauge group (either $E_8\times
E_8$ or $SO(32)$). From the four dimensional viewpoint the moduli
$T_{IJ}$ and $Y_{I}^{k}$  correspond to  the vev's of massless scalar
fields, members of the $N=4$  vector supermultiplets.

The explicit form of the $N=4$ heterotic partition function
$\Gamma^{SO(32)}_{(6,22)}[T_{IJ},Y_{I}^{k}]$ is:

$$
\Gamma^{SO(32)}_{(6,22)}(T,Y)(\tau,{\bar \tau})={(det
G)^{3}\over \tau_2^3}\sum_{m^I,n^I}~
\exp\left[-\pi~T_{IJ}\frac{ (m^I+\tau n^I)(m^J+{\bar \tau}n^J)
}{\tau_2}
\right]
$$
\be
\times~\frac{1}{2}~\sum_{\gamma,\delta}
\prod_{k=1}^{16}~\exp\left[~{i\pi\over
4}(n^I~Y^k_I~Y^k_J~m^J+2~\delta~Y^k_I~n^I)\right]~
{\bar\th}
\left[^{\gamma +n^I~Y_I^k}_{\delta+m^I~Y_I^k}\right]({\bar \tau})
\label{b3}\ee

When all $Y$--moduli are zero ($Y_{I}^{k}=0$) then the gauge group is
extended from
$G=U(1)^{22}$ to   $G=U(1)^6\times SO(32)$.

An alternative representation of $\Gamma_{(6,22)}[T_{IJ},Y_{I}^{k}]$
is the one in which, for $Y_{I}^{k}=0$, the extended gauge symmetry is
$G=U(1)^6\times E_8 \times E_8$ instead of $U(1)^6 \times SO(32)$:

$$
\Gamma^{E_8 \times E_8}_{(6,22)}(T,Y)(\tau,{\bar \tau})={(det
G)^{3}\over \tau_2^3}\sum_{m^I,n^I}~
\exp\left[-\pi~T_{IJ}\frac{ (m^I+\tau n^I)(m^J+{\bar \tau}n^J)
}{\tau_2}
\right]
$$
\be
\times
{}~\frac{1}{2}~\sum_{\gamma_1,\delta_1}\prod_{k=1}^{8}
\exp\left[~{i\pi\over 4}(n^I~Y^k_I~Y^k_J~m^J+2~
\delta_1~Y^k_I~n^I)\right]~
{\bar\th}
\left[^{\gamma_1+n^I~Y_I^k}_{\delta_1+m^I~Y_I^k}\right]({\bar \tau})
\label{b4}\ee
$$
\times
\frac{1}{2}~\sum_{\gamma_2,\delta_2}\prod_{k=9}^{16}
\exp\left[~{i\pi\over 4}(n^I~Y^k_I~Y^k_J~m^J+2~
\delta_2~Y^k_I~n^I)\right]~
{\bar\th}
\left[^{\gamma_2+n^I~Y_I^k}_{\delta_2+m^I~Y_I^k}\right]({\bar \tau})
$$

Both the $SO(32)$ and $E_8\times E_8$ representations are connected
continuously by  marginal deformations with  the $G=SO(44)$ maximal
gauge symmetry  point:
\def\wb{\bar W}
\be
\Gamma^{SO(44)}_{(6,22)}(\tau,{\bar \tau})=
{}~\frac{1}{2}~\sum_{\gamma,\delta}~\th^6
\left[^{\gamma}_{\delta}\right](\tau)
{\bar \th}^{22}\left[^{\gamma }_{\delta}\right]({\bar \tau})
\label{b5}\ee

Another useful representation of the $\Gamma_{(6,22)}$ is that of the
lorenzian
left- and right-momentum even self-dual lattice. This representation is
obtained by performing Poisson resummation on $m^I$ using either
$\Gamma^{SO(32)}_{(6,22)}(T,Y)$ or $\Gamma^{E_8 \times
E_8)}_{(6,22)}(T,Y)$:
\be
\Gamma_{(6,22)}(P_I,{\bar P}_I, Q^k)= \sum_{P_I,{\bar P}_I,
Q^k}~\exp~\left[{i\pi\tau\over 2}P_I~G^{IJ}P_J-
{i\pi{\bar \tau}\over 2}{\bar  P}_IG^{IJ}{\bar P}_J
-i\pi{\bar \tau }{\hat Q}^k{\hat Q}^k\right]
\label{b6}\ee
with
\be
{1\over 2}~P_I~G^{IJ}~P_J~-~{1\over 2}~{\bar  P}_I~G^{IJ}~{\bar
P}_J~-~{\hat Q}^k{\hat Q}^k~=~{\rm even~~integer}
\label{b7}\ee

In the above equations $G^{IJ}$ is the inverse of $G_{IJ}$; the lattice
momenta $P_I,~{\bar P}_I$, and the left charges ${\hat Q}^k$ are qiven
in terms of
the moduli parameters $G_{IJ}, B_{IJ}$ and $Y^k_I$ and in terms of the
charges
($m_I,~n^I,~Q^k$):
$$
P_I~=~
m_I~+~Y_I^k~Q^k~+~{1\over2}Y_I^k~Y^k_J~n^J~+B_{IJ}~n^J~+~G_{IJ}~n^J
$$
\be
{\bar P}_I~=~
m_I~+~Y_I^k~Q^k~+~{1\over2}Y_I^k~Y^k_J~n^J~+B_{IJ}~n^J~-~G_{IJ}~n^J
\label{a1}\ee
$$
{\hat Q^k}~=~Q^k~+~Y_I^k~n^J
$$
All N=4 heterotic solutions are defined in terms of the vacuum
expectation values of the moduli fields $(T_{IJ},~Y_I^k)$ and thus
different solutions are connected to each other by a
string-Higgs phenomenon. At some special points of the moduli space, we
have extensions of the gauge group as in the effective $N=4$
supergravity theories. In string theories, a further extension can
occur due to the non-zero winding charges ($n^I$) which can become
massless in special points of the moduli space. Thus, in string theory,
a large class of disconnected $N=4$ supergravities are continuously
related to one another due to the existence of the winding states.
This precise fact is the origin of the
perturbative string unification of all interactions in string theories.

There is another way of writing the left and right conformal weights
for a (6,22) lattice.
Introduce the $28\times 28$ matrices
\be
L=\left(\matrix{0&{\bf 1}_{6}&0\cr
{\bf 1}_{6}&0&0\cr
0&0&-{\bf 1}_{16}\cr}\right)
\label{a7}\ee
which is the $O(6,22)$ invariant metric, and
\be
M=\left(\matrix{G^{-1}& G^{-1}C &G^{-1}Y^{t}\cr
C^{t}G^{-1}&G+C^{t}G^{-1}C+Y^{t}Y&C^{t}G^{-1}Y^{t}+Y^{t}\cr
YG^{-1}&YG^{-1}C+Y&{\bf 1}+YG^{-1}Y^{t}\cr}\right)
\label{a8}\ee
with
\be
C_{IJ}=B_{IJ}+\sum_k {1\over 2}Y^{k}_{I}Y^{k}_{J}
\label{a9}\ee
Notice that  $M$ is a symmetric element of $O(6,22)$
so that $M^{t}LM=MLM=L$, and that  its inverse reads
\be
M^{-1}= LML = \left(\matrix{G+C^{t}G^{-1}C+Y^{t}Y&
C^{t}G^{-1}&-(C^{t}G^{-1}+{\bf 1})Y^{t}\cr
G^{-1}C& G^{-1}& -G^{-1}Y^{t}\cr
-Y(G^{-1}C+{\bf 1})& -YG^{-1}& {\bf 1}+YG^{-1}Y^{t}\cr}\right)
\label{a10}\ee
Introduce also the 28-vector of charges
\be
\vec a \equiv (\vec m,\vec n,\vec Q)
\label{aa10}\ee
Then
\be
p_L^2\equiv P_I G^{IJ} P_J= a_i (M+L)_{ij} a_j\;\;\;,\;\;\;P_R^2
\equiv \bar P_I G^{IJ} \bar P_J+2\hat Q^k\hat Q^k= a_i (M-L)_{ij} a_j
\label{aa11}\ee
Thus
\be
\Gamma_{(6,22)}=\sum_{\vec a\in {\cal E}_{6,22}}\;q^{{1\over
4}a^T(M+L)a}
\bar q^{{1\over 4}a^T(M-L)a}
\label{bb1}\ee
where the lattice ${\cal E}_{6,22}$ is the even self-dual lattice
${\cal E}_{1,1}^6\otimes {\cal E}_{16}$ and ${\cal E}_{16}$ is the
Spin(32)/Z$_2$ lattice.
${\cal E}_16$ can be described by the roots and spinor weight of
SO(32).
Let $u_i$ $i=1,2,\cdots,16$ be an orthonormal basis, $u_i\cdot
u_j=\delta_{ij}$.
$u_i$ is a 16-dimensional vector with 1 in the $i$-th entry and 0
elsewhere.
The SO(32) roots are:
\be
f_i=u_i-u_{i+1}\;\;\;,\;\;\;i=1,2,...,15\;\;\;,\;\;\;
f_{16}=u_{15}+u_{16}
\label{bb2}\ee
The spinor weight is $f_s=-(\sum_{i=1}^{16}u_i)/2$.
We have
\be
f_i\cdot f_j=2\delta_{ij}-\delta_{i,j+1}-\delta_{i+1,j}
\label{bb4}\ee
\be
f_i\cdot f_s=-\delta_{i,16}\;\;\;,\;\;\;f_s\cdot f_s=2
\label{bb5}\ee
Then an arbitrary lattice vector can be written as
\be
l=\sum_{i=1}^{16}n_i\;(f_i+\alpha f_s),\;\;\;,\;\;\;\alpha=0,1
\label{bb3}\ee

The low energy effective N=4 supergravity is  manifestly
 invariant under the
full
O(6,22) group, acting on the fields in the following way
\be
M\to \Omega\;M\Omega^T\;\;\;,\;\;\;F_{\mu\nu}\to\Omega\;F_{\mu\nu}
\label{a11}\ee
where $\Omega\in$ O(6,22), i.e. $\Omega^T \Omega=I$.
The full string theory is invariant under the discrete subgroub
O(6,22,Z),
and the (electric) charges transform as
\be
a_i\to \Omega_{ij}a_j
\label{aa12}\ee

Furthermore the equations of motion and Bianchi identities are
left invariant by  the $SL(2,R)$ transformation
\be
S \to {aS+b\over cS+d}\; \;\;,\;\; M\to M\;\;\;,\; \; F_{\mu\nu}^i\to
(c\;{\rm Re}S + d) F_{\mu\nu}^i +
 c\;{\rm Im}S\; (ML)_{ij} \ ^*F_{\mu\nu}^j
\label{a12}\ee
with $ad-bc=1$.
In particular, the transformation $S\to -1/S$ interchanges electric and
magnetic charges.
It has been conjectured  \cite{S}--\cite{ss},
that a discrete subgroup $SL(2,Z)$ of this continuous symmetry of the
equations of motion of the effective theory is a (non-perturbative)
symmetry of the full theory.
For this to be true we will have to include in the theory states that
carry
both electric and magnetic charges. Magnetic or dyonic states are
non-perturbative since the full perturbative heterotic spectrum is
electrically charged.

Following references \cite{sen,cv}--\cite{CCLMR}
let us parametrize the electric
and magnetic charges in terms of the integer-valued 28-vectors
$\vec\a,\vec\b$  and the moduli
as follows:
\be
\vec Q_{e}={1\over \sqrt{2}\; {\rm Im}S}M (\vec\a+{\rm Re}S
\;\vec\b)\;\;\;,\;\;\;\vec Q_m = {1\over\sqrt{2}} L\;\vec\b
\label{a13}\ee
This parametrization incorporates automatically the
 Dirac-Schwinger-Zwanziger-Witten quantization condition for dyons
with a $\theta$-angle.
The BPS mass formula can then
 be expressed in two equivalent ways
\def\ub{\bar U}
\def\tb{\bar T}
\be
 M_{BPS}^2= {{\rm Im}S\over 4}\; \left[
Q_e^t {\tilde M}_+  Q_e + Q_m^t  {\tilde M}_+  Q_m +2 \sqrt{
(Q_e^t  {\tilde M}_+  Q_e)(Q_m^t {\tilde M}_+  Q_m) -
(Q_e^t  {\tilde M}_+   Q_m)^2}\;
\right]
\label{a14}\ee
$$
={1\over 4 \im} (\a^t+S\b^t)  M_{+} (\a+\bar
S\b)+ {1\over 2}\;\sqrt{(\a^t M_+\a)(\b^t M_+\b)-(\a^t M_+\b)^2}
$$

\noindent with $M_+=M+L$ and $ {\tilde M}_+ = LM_+L$.
The square-root factor in the above expressions
 is proportional to  the difference of the two
central charges squared: depending on whether  this vanishes
or not, the representation
preserves  1/2  or  1/4 of the supersymmetries, and is thus
either short or intermediate.
For perturbative BPS
 states of the heterotic string,
$\vec\b = 0$. Thus they belong to short multiplets.
 Their  mass  reads

\def\im{{\rm Im}S}
\be
M^2_{BPS, {\rm pert}}={1\over 4\ \im }\;\a^t M_+\a=
{1\over
4\ \im }\;p^2_{L}
\label{a15}\ee
The factor of $\im$ is there because masses are
 measured in units of $M_{\rm Planck}$.

\vskip 1cm

The BPS mass formula is manifestly invariant under
$O(6,22,Z)$ acting on the fields as
in (\ref{a11}) and on the charge vectors as
\be
\a\;\to\; \Omega\;\a\;\;\;,\;\;\;\b\;\to\; \Omega\;\b\ ,
\label{a16}\ee
It is also invariant  under
$SL(2,Z)_S$ acting on the fields as in (\ref{a12}) and
on the vectors as
\be
\left(\matrix{\vec\a\cr\vec\b\cr}\right) \;\to\;
\left(\matrix{a&-b\cr-c&d\cr}\right)
\left(\matrix{\vec\a\cr\vec\b\cr}\right)
\label{a17}\ee

It can be checked that (\ref{a14}) is
$SL(2,Z)_S$
invariant.
The spectrum of N=4 heterotic string
theory is
mapped to that of N=4 type II string theory under a
transformation
which is  an avatar of
string-string
duality in 6-d \cite{HT}. The two low energy field theories are
generically
distinct, but they coincide when the R-R gauge fields are set to
zero.

We will analyse the BPS mass formula in a subspace of the full (6,22)
moduli
space. In particular we will keep the 4 real moduli of a 2-torus as
well as
the 16 Wilson-line moduli $Y^i_\a$, $i=9,2,\cdots,24$,
$\a=1,2$, associated with the 2-torus.
This is a subspace of the full moduli space with an O(2,18) structure
and it
will be relevant for the models where N=4 is spontaneously broken to
N=2.
We parametrize
\be
G={{T_{2}-{1\over 2U_2}\sum_{i} ({\rm Im}W_{i})^2}\over
U_2}\left(\matrix{1&U_1\cr
U_1&|U|^2\cr}\right)\;,\;
B=\left(T_1-{\sum_{i}{\rm Re}W_{i}{\rm Im}W_{i}\over
2U_2}\right)\left(\matrix{0&1\cr -1&0\cr}\right)
\label{a26}\ee
\be
W_i=-Y^i_{2}+ U Y^i_{1}
\label{a27}\ee
Define $e^{-K}=T_2U_2-{1\over 2}\sum_i{\rm Im}W_i^2$.
$K$ is the K\"ahler potential for the moduli.

The perturbative part (electric charges only) becomes the well-known
SO(2,18) invariant mass formula
\be
M^2_{\rm pert}={1\over 4 \;S_2
\left(T_2U_2-{1\over 2}\sum_i{\rm
Im}W_i^2\right)}|-m_1U+m_2+Tn_1+(TU-{1\over 2}\sum_i
W_i^2)n_2+W_iq^i|^2
\label{am36}\ee

The quantity under the square root in (\ref{a14}) is a
perfect square,
\be
\sqrt{(\a\cdot M_+\cdot\a)(\b\cdot M_+\cdot\b)-(\a\cdot
M_+\cdot\b)^2}=
|\a\cdot F\cdot \b|
\label{a20}\ee
Then (we assume for the moment that the number in the absolute value is
positive),
\be
M_++i\; F={e^K}\; R
\label{a28}\ee
where $R$ is the following complex matrix
\be
R=\left(\matrix{R_{11}&R_{12}\cr R_{21}&R_{22}\cr}\right)
\label{a29}\ee
$R_{11}$ is a $4\times 4$ matrix given by
\be
R_{11}=\left(\matrix{|U|^2&-U&-U\bar T&U\left({1\over
2}\sum_i\bar W_i^2-\bar T\bar U\right)\cr
-\bar U&1 & \bar T &\bar T\bar U-{1\over 2}\sum_i\bar W_i^2\cr
-T\bar U&T& |T|^2& T\left(\bar T\bar U-{1\over 2}\sum_i\bar
W_i^2\right)\cr
\bar U\left({1\over 2}\sum_i W_i^2-TU\right)&
TU-{1\over 2}\sum_i W_i^2&\bar T\left(TU-{1\over 2}\sum_i W_i^2
\right)&|TU-{1\over 2}\sum_i W_i^2|^2\cr}\right)
\label{a30}\ee

$R_{12}$ is a $4\times 24$ matrix,
\be
R_{12}=\left(\matrix{...&- U \bar W_i&...\cr
...&\bar W_i&...\cr
...&T\bar W_i&...\cr
...&\left(TU-{1\over 2}\sum_i W_i^2\right)\bar W_i&...\cr}\right)
\label{a31}\ee
while
\be
R_{21}=\overline{R^{t}_{12}}
\label{a32}\ee
and finally $R_{22}$ is a $24\times 24$ matrix
\be
R_{22,ij}=W_{i}\bar W_{j}
\label{a33}\ee
The matrix $R$ satisfies $R^{t}=\bar R$ which implies (as it should)
that
$M_+$ is symmetric and $F$ is antisymmetric.

Then the BPS formula can be cast in the form (when $\a \cdot F\cdot
\b$ is positive)
\be
M^2_{BPS,+}={(\a+S\b)\cdot R\cdot (a+\bar S\b)\over
4\;S_2\left(T_2U_2-{1\over 2}\sum_i{\rm Im}W_i^2\right)}
\label{a34}\ee
The matrix $R$ has also the property:
\be
R=\left(\matrix{-U,&1,&T,&TU-{1\over 2}\sum_i
W_i^2,&W_i\cr}\right)\otimes
\left(\matrix{-\ub\cr 1\cr \tb\cr \tb\ub-{1\over 2}\sum_i \wb_i^2\cr
\wb_i\cr}\right)
\label{a35}\ee
which implies that det$R=0$ and
$$
M^2_{BPS,+}={1\over 4 \;S_2
\left(T_2U_2-{1\over 2}\sum_i{\rm
Im}W_i^2\right)}|-m_1U+m_2+Tn_1+(TU-{1\over 2}\sum_i
W_i^2)n_2+W_iq^i+
$$
\be
+S[-\tilde
m_1U+\tilde m_2+T\tilde
n_1+
\tilde n_2(TU-{1\over 2}\sum_i W_i^2)+\tilde q^i W_i]|^2
\label{a36}\ee
Again, in the negative case $S\to \bar S$.

Now the $O(2,18,Z)$ (T-duality) transformations are:
\be
T\to T+1\;\;\;,\;\;\;W_i\to W_i
\label{a37a}\ee
\be
U\to U+1\;\;\;,\;\;\;W_i\to W_i
\label{a37b}\ee
\be
T\leftrightarrow U\;\;\;,\;\;\;W_i\to W_i
\label{a37c}\ee
\be
T\to -{1\over T}\;\;\;,\;\;\;U\to U-\sum_i{W^2_i\over
2T}\;\;\;,\;\;\;W_{i}\to {W_{i}\over T}
\label{a37d}\ee
\be
U\to -{1\over U}\;\;\;,\;\;\;T\to T-\sum_i{W^2_i\over
2U}\;\;\;,\;\;\;W_{i}\to {W_{i}\over U}
\label{a37e}\ee

\be
W_{i}\to W_{i}+a_i U\;\;,\;\;U\to U\;\;,\;\;T\to T+a_iW_i+{1\over
2}\sum_{i}a_{i}^{2}~U
\label{a37f}\ee

\be
W_{i}\to W_{i}+a_i\;\;,\;\;U\to U\;\;,\;\;T\to T
\label{a37g}\ee

The mass formula here is not invariant under heterotic-type II
duality.
The reason is that the Wilson lines we turned on correspond to R-R
gauge fields in the type II side. The mass formula in this case is
mapped to the type II BPS formula.

To go further we will set $W_i=0$.
We  obtain
\be
M^2_{BPS,+}={|-m_1U+m_2+T(n_1+n_2U)+S[-\tilde m_1U+\tilde
m_2+T(\tilde
n_1+
\tilde n_2U)]|^2\over 4\;U_2T_2S_2}
\label{a24}\ee
$$
={|-m_1U+m_2+T(n_1+n_2U)+S(-\tilde m_1U+\tilde m_2)+ST(\tilde n_1+
\tilde n_2U)|^2\over 4\;U_2T_2S_2}
$$
Cast in this form, (\ref{a24}) is invariant under $S\leftrightarrow T$
interchange \footnote{This was first observed at the level of the
action in
\cite{duff}. More indications for the string-string duality
conjecture were provided in \cite{HT}.} :
\be
T\leftrightarrow S\;\;\;,\;\;\;\left(\matrix{m_1\cr m_2\cr n_1 \cr
n_2\cr}\right)
\to\left(\matrix{m_1\cr m_2\cr \tilde m_2 \cr-\tilde
m_1\cr}\right)\;\;,\;\;
\left(\matrix{\tilde m_1\cr \tilde m_2\cr \tilde n_1 \cr \tilde
n_2\cr}\right)
\to\left(\matrix{-n_2 \cr n_1\cr \tilde n_1\cr \tilde n_2\cr}\right)
\label{a24a}\ee
This is precisely the transformation implied by string-string duality
at the
action level, as we will see in detail in another section.
Note also that for generic moduli the heterotic-type II duality
transformation on electric and magnetic charges is still given by
(\ref{a24a}), all other charges being invariant, while the map on the
moduli themselves becomes more complicated.

If the quantity in the absolute value is negative we  obtain
(\ref{a24})
with $S\to \bar S$
\be
M^2_{BPS,-}={|-m_1U+m_2+T(n_1+n_2U)+\bar S[-\tilde m_1U+\tilde
m_2+T(\tilde n_1+
\tilde n_2U)]|^2\over U_2T_2S_2}
\label{a25}\ee
$$={|m_1U-m_2-T(n_1+n_2U)-\bar S[\tilde m_1U-\tilde m_2-T(\tilde
n_1+
\tilde n_2U)]|^2\over U_2T_2S_2}
$$
This can be obtained from (\ref{a24}) by changing the signs of
electric and magnetic charges as well as $S_1$.
This formula looks somewhat peculiar. We will see however that for Type
II
perturbative states that break 3/4 of the supersymmetry, it correctly
describes their spectrum.
Also (\ref{a25}) is $SL(2,Z)_S$ invariant,  but also invariant now
under
$T\leftrightarrow \bar S$ interchange.\footnote{Similar observations
were
made in \cite{CCLMR}.}

Although the mass formula for non-perturbative BPS states is understood
we do not know a priori the multiplicities of all these states.
{}From the $N=4$ heterotic string we know the multiplicities when
$\beta_i=0$. Using $SL(2,Z)$ we also know the multiplicities of all
states with
$\alpha\cdot\beta~=~0$. To go further and learn more about the states
with
$\alpha\cdot\beta~\ne ~0$ (namely intermediate multiplets) it is
necessary
to go beyond the string picture and learn more about the
non-perturbative structure of the theory.
The heterotic string on $T^6$ is suposed to be equivalent in the strong
coupling limit with the type II theory compactified on $K_3\times T^2$.
Moreover, there is a hypothetical 11-d theory (M-theory) that includes
the
non-perturbative dynamics of type IIA theory \cite{wi}.
Thus compactification of M-theory on $K_3\times T^3$ contains all the
relevant non-perturbative information about the heterotic N=4 theory.
This idea led to a  conjecture on the multiplicities of dyonic BPS
states in the 4-d N=4 theory \cite{dvv}.
This will be an important input, for our non-perturbative analysis of
the spontaneously broken N=4 theory.

\section{$N=4 \rightarrow N=2$ partial spontaneous breaking of
supersymmetry }
\setcounter{equation}{0}

One of the defining characteristics of the $N=4$ theories is that the
states are classified by their transformation properties under the
$R$-symmetry group which, for N=4 supersymmetry, is $G_R$=SU(4)$\sim$
SO(6).
In the gravitational multiplet the gravitinos are in ${\bf 4}$
representation of $G_R$, the graviphotons are in ${\bf 6}$,  while
the graviton, the dilaton and the antisymmetric tensor field  are
singlets. The degrees of freedom of a massless $N=4$ vector multiplet
are also in definite representations of $G_R$: the scalars are in
${\bf 6}$, the gauginos
are in ${\bf 4}$, while the gauge bosons are singlets. In the
heterotic string, $G_R$
is constructed in terms of the six left-moving compactified
supercoordinates, ($\Phi^I,~\Psi^I$). The world-sheet fermion
bilinears $\Psi^I~\Psi^J$ form an $SO(6)_{k=1}$ Kac--Moody algebra.
In the light-cone picture, the full spectrum of the theory is
classified in representations of $SO(6)_{k=1}$ and in terms of the
$U(1)_{0}$ helicity charge $q^0=\oint j^0$,
$~j^0=\Psi^{\mu}\Psi^{\nu}$, $\mu,\nu=3,4$. In the $N=4$ spectrum the
three internal helicity charges $q^i=\oint j^i$,
$~j^i=\Psi^{2k-1}\Psi^{2k}$, $k=1,2,3$ and $q^0$ are all
simultaneously integers for space-time bosonic states and
simultaneously half-integers for the fermionic states:
\vskip .2cm
$q^i=$ half-integers for spacetime fermions
\vskip .2cm
$q^i=$ integers for spacetime bosons.
\be ~ \label{bb6}\ee
 Furthermore all physical states have odd total $q^i$ charge
(GSO-projection)
\be
q^0~+~q^1~+q^2~+~q^3~=~{\rm odd~~integer}.
\label{bb7}\ee
The last condition remains valid for supersymmetric solutions with
less than four supersymmetries. In order to have a lower number of
supersymmetries, the $q^i$'s
must not be simultaneously integers or half-integers. It is then
necessary to
modify the world-sheet action $S^{2d}$, adding background fields that
can change
the individual values of the $q^i$'s, keeping however their total
$q^{i}$ charge:
\be
\Delta S^{2d}=\int dzd{\bar z}~F_{IJ}^a~(\Psi^I~\Psi^J~-~\Phi^I
\lrp ~\Phi^J)~{\bar J}^a,
\label{b8}\ee
where ${\bar J}^a$ denotes any dimension (0,1) operator. The part of
the left-moving operator $(\Phi^I\lrp \Phi^J)$ is necessary to
ensure the $N=(1,0)$ super-reparametrization  of the 2-d action. From
a higher-dimensional point of view, the $F^a_{IJ}$ denote
non-trivial gauge or gravitational (${\cal R}^{(KL)}_{IJ}$) field
backgrounds. In four dimensions they give rise to  non-vanishing
auxiliary fields. The permitted values of $F^a_{IJ}$ (${\cal
R}^{(KL)}_{IJ}$) are not arbitrary. Only those
for which
\be
U_L(F)=\exp~[\int dz~
F_{IJ}^a~(\Psi^I~\Psi^J~-~\Phi^I\lrp~\Phi^J)]
\label{b9}\ee
commutes with the 2-d supercurrent (~$T_F=\Psi^{\mu}~\partial
\Phi^{\mu}+\Psi^I~\partial \Phi^I$) are allowed. This restriction
generates   a quantization of the permitted $F^a_{IJ}$ (${\cal
R}^{(KL)}_{IJ}$) backgrounds.

A partial $N=4~\rightarrow ~N=2$ breaking is possible \cite{ss1} when
$F^a_{3,4}=-F^a_{5,6}= H$ is not zero (self-duality condition).
Indeed, in that case the $q^2$ and $q^3$ charges are shifted,
preserving the total $q^i$ charge. In order to define the full
deformation of the spectrum it is necessary to find a representation
of the partition function in which  the bosonic charges
\be
Q^B_2=\oint dz \Phi^3\lrp \Phi^4
{}~{\rm and}~
Q^B_3=\oint dz \Phi^5\lrp \Phi^6
\label{b10}\ee
are well defined. As a starting point we fermionize the four internal
bosonic coordinates
\be
\partial \Phi^I=y^Iw^I ~{\rm and}~ {\bar \partial} {\bar
\Phi}^I={\bar y}^I~{\bar w}^I, ~I=3,4,5,6.
\label{b11}\ee
In this representation the two dimensional  supercurrent is
\cite{abk},\
\be
T_F=\Psi^{\mu}~\partial \Phi^{\mu}~+\sum_{I=1}^2 \Psi^I~\partial
\Phi^I~+\sum_{I=3}^6 \Psi^I~y^I~w^I.
\label{b12}\ee
We will now perform the following $Z_4$ transformation:
$$
\Psi^3 \rightarrow ~~\Psi^4, ~~~ y^3 \rightarrow ~~y^4,~~~\Psi^5
\rightarrow -\Psi^6, ~~~ y^5 \rightarrow -y^6,
$$
\be
\Psi^4 \rightarrow -\Psi^3, ~~~ y^4 \rightarrow -y^3,~~~~~\Psi^6
\rightarrow ~~\Psi^5, ~~~ y^6\rightarrow ~~y^5,
\label{b13}\ee
$$
w^3\rightarrow ~w^4,~~~~w^4\rightarrow ~w^3,~~~~~~w^5\rightarrow
{}~w^6,~~~~~~w^6\rightarrow ~w^5,
$$
which leaves (\ref{b12}) invariant.
The  above transformation corresponds to a $\pi /2$ rotation on the
complex fermion basis:
\be
\chi_1 \rightarrow ~ e^{2i\pi\phi}~\chi_1\;\;\;,\;\;\;
\chi_2 \rightarrow ~ e^{-2i\pi\phi}~\chi_2\;\;\;,\;\;\;
Y_1 \rightarrow ~ e^{2i\pi\phi}~Y_1\;\;\;,\;\;\;Y_2 \rightarrow ~
e^{-2i\pi\phi}~Y_2
\ee
\be
W_{+} \rightarrow W_{+}\;\;\;,\;\;\;W_{-} \rightarrow  e^{4i\pi
\phi}W_{-}
\ee
where
\be
\chi_1={{\Psi^3~+~i\Psi^4}\over
\sqrt{2}}\;\;\;,\;\;\;\chi_2={{\Psi^5~+~i\Psi^6}\over
\sqrt{2}}\;\;\;,\;\;\;Y_1={{y^3~+~iy^4}\over
\sqrt{2}}\;\;\;,\;\;\;Y_2={{y^5~+~iy^6}\over \sqrt{2}}
\ee
\be
W_{+}={{(w^3+w^4)+i(w^5+w^6)}\over {2}}\;\;\;,\;\;\;
W_{-}={{(w^3-w^4)+i(w^5-w^6)}\over {2}}
\ee
Similarly for the right-moving degrees of freedom (${\bar \Psi}^I$,
${\bar y}^I, ~{\bar w}^I,~I=3,4,5,6$).
The above transformation is a symmetry only if the rotation angle is
a multiple of $\pi/2$ or  $\phi~=k/4$, with $k$ integer.

Observe that with the help of the world-sheet fermions we  can
classify the $N=4$ string spectrum  in terms  of a left and a right
$U(1)$ charges
$Q_L=\oint j_L$ and $Q_R=\oint j_R$, where
$$
j_L=\chi_1\chi^{\dagger}_1-\chi_2\chi^{\dagger}_2+
Y_1Y^{\dagger}_1-Y_2Y^{\dagger}_2
+2W_{-}W^{\dagger}_{-},~~~~~~~
$$
\be
Q_L=q_{\chi_1}-q_{\chi_2} + q_{Y_1}-q_{Y_2}+2q_{W_-}
\label{b15}\ee
and
$$
j_R={\bar \chi}_1{\bar \chi}^{\dagger}_1-
{\bar \chi}_2{\bar
\chi}^{\dagger}_2+{\bar Y}_1{\bar Y}^{\dagger}_1-
{\bar Y}_2 {\bar Y}^{\dagger}_2
+2{\bar W}_{-} {\bar W}^{\dagger}_{-},~~~~~~~~~
$$
\be
Q_R=q_{{\bar
\chi}_1}-q_{{\bar \chi}_2}+{\bar q}_{Y_1}-
{\bar q}_{ Y_2}+2{\bar
q}_{W_-}
\label{b16}\ee

We are now in a position to switch on non-vanishing $F_{IJ}^a$ by
performing a boost among the fermionic charge lattice  and  the
$\Gamma_{(2,n)}$ lattice:
$$
q_{\chi_1}\rightarrow q_{\chi_1}+h_in^i\;\;\;,\;\;\;
q_{\chi_2}\rightarrow q_{\chi_2}-h_i n^i\;\;\;,\;\;\;q_{{\bar \chi}_1}
\rightarrow q_{{\bar \chi}_1}+h_in^i\;\;\;,\;\;\;
q_{{\bar \chi}_2}\rightarrow q_{{\bar
\chi}_2}-h_i n^i
$$
$$
q_{Y_1}\rightarrow q_{Y_1}+h_in^i,
\;\;\;,\;\;\;q_{Y_2}\rightarrow q_{Y_2}-h_in^i\;\;\;,\;\;\;q_{{\bar
Y}_1}\rightarrow q_{{\bar Y}_1}+h_in^i,
\;\;\;,\;\;\;q_{{\bar Y}_2}\rightarrow q_{{\bar Y}_2}-h_in^i
$$
\be
q_{W_-} \rightarrow q_{W_-}+2h_in^i\;\;\;,\;\;\;q_{W_+} \rightarrow
q_{W_+}
\;\;\;,\;\;\;q_{{\overline W}_-}\rightarrow q_{{\overline
W}_-}+2h_in^i,
\;\;\;,\;\;\;q_{{\overline W}_+}\rightarrow q_{{\overline W}_+}
\label{b17}\ee
$$
P^L_i(h_i)=P^L_i-h_i(Q_L-Q_R,)\;\;\;,\;\;\;
P^R_i(h_i)=P^R_i-h_i(Q_L -Q_R)
$$
with
$$
P^L_i=m_i~+~Y_i^a~Q^a~+~{1\over2}Y_i^a~Y^a_j~n^j~
+B_{ij}~n^j~+~G_{ij}~
n^j
$$
$$
P^R_i=m_i~+~Y_i^a~Q^a~+~{1\over2}Y_i^a~Y^a_j~n^j~
+B_{ij}~n^j~-~G_{ij}~
n^j
$$
\be ~~~~\ee
$Y^a_i$ $i=i,2$, $a=1,2,...,18$ are the Wilson-line moduli of the
$\Gamma (2,18)$ lattice.

Owing to the non-zero $h_i$ shift, two of the $N=4$ gravitini
become massive, with mass proportional to $|q_{\chi_1}-q_{\chi_2}|$.
The $N=4$ gravitini
have vanishing $m_i, n^i, q_{Y_i}, q_{{\bar Y}_i}, q_{W_{i}},
q_{{\bar W}_{i}}$ charges. The two of them remain massless since
$|q_{\chi_1}-q_{\chi_2}|=0$, while the other two   become massive
since $|q_{\chi_1}-q_{\chi_2}|=1$:
\be
(m^2_{3/2})_{1,2}=0, ~~~(m^2_{3/2})_{3,4}={|F|^2\over 4{\rm
Im}T{\rm Im}U}, ~~~
\label{b18}\ee
${\rm with}~F~=~h_1~+U~h_2$, $T$ and $U$ being the usual complex
moduli of the $\Gamma_{(2,2)}$ lattice.

The global existence of the supercurrent implies in this case the
quantization condition:
$4h_i=$ integer. The $N=2$ partition function $\Z^{4\rightarrow
2}(F)$ is obtained from that of $N=4$ by shifting  the lattice
momenta $P_i$ and the $R$-charges $q_i$ as above. Performing a
Poisson resummation on $m_i$, we obtain the following expression:
\vskip .2cm

\centerline{ $ \gamma = 2h_i n^i.~~~\delta = 2h_i m^i,~~~F=h_1+U~h_2$
}

$$
\Z^{4\rightarrow 2}(F)= {(\tau_2)^{-1} \over \eta^2{\bar
\eta}^2}\sum_{m^i,n^i}
\sum_{\alpha, \beta}{1\over 2}(-)^{\alpha+\beta +\alpha \beta}
\frac{\th^{2}[^{\alpha}_{\beta}]}{\eta^2}\frac{\th[^{\alpha~+~
\gamma}_{\beta~+~\delta}]}{\eta}
\frac{\th[^{\alpha~-~\gamma}_{\beta~-~\delta}]}{\eta}
{}~{\Gamma_{(2,2)}[^{n^i}_{m^i}]\over
|\eta|^4}~Z_{(4,4)}[^{\gamma}_{\delta}]\times
$$
\be
\times~\sum_{{\bar \alpha}, {\bar \beta}}~
\frac{1}{2}~\sum_{\bar \alpha, \bar \beta}
\frac{{\bar \th}^{6}[^{\bar \alpha}_{\bar \beta}]}{{\bar \eta}^6}
{}~\frac{{\bar \th}[^{{\bar \alpha}~+~{\bar \gamma}}_{{\bar
\beta}~+~{\bar \delta}}]}{{\bar \eta}}~\frac{{\bar \th}[^{{\bar
\alpha}~-~{\bar \gamma}}_{{\bar \beta}~-~{\bar \delta}}]}{{\bar
\eta}}
{}~\sum_{\epsilon, \zeta}\frac{1}{2}~\frac{{\bar
\th}^{8}[^{\epsilon}_{\zeta}]}{{\bar \eta}^8},
\label{b19}\ee
where
\be
\Gamma_{(2,2)}[^{n^i}_{m^i}]={\sqrt {{\rm det}~G}\over \tau_2}
{\rm exp}~\left[-\pi G_{ij} \frac{(m^i+n^i\tau)(m^j+n^j~{\bar
\tau})}{\tau_2}+2i \pi B_{ij}m^in^j\right]
\ee
and
\be
Z_{(4,4)}[^{\gamma}_{\delta}]=
\frac{1}{2}\sum_{a,b} \left|\frac{\th[^{a}_{b}]}{\eta}
\frac{\th[^{a+2\gamma}_{b+2\delta}]}{\eta}
\frac{\th[^{a+\gamma}_{b+\delta}]}{\eta}
\frac{\th[^{a-\gamma}_{b-\delta}]}{\eta}\right|^2
\label{b20}\ee
When $h_i=0$  $(\gamma,~\delta=0)$,  $\Z^{4\rightarrow 2}(F=0)$
corresponds to
the $N=4$ heterotic string solution based on a gauge group
$U(1)\times U(1) \times SO(8) \times E_8 \times E_8$;  the $SO(8)$
gauge group factor corresponds to the extended symmetry of the
$\Gamma_{(4,4)}$ lattice at the fermionic point
\be
Z_{(4,4)}[^0_0]=\frac{1}{2}~\sum_{a,b}~
\left|\frac{\th[^{a}_{b}]}{\eta}\right|^8
\label{b21}\ee
The sum over $m^i$ and $n^i$ gives rise to the $\Gamma (2,2)$ lattice
at an arbitrary point of the moduli space:
\be
\sum_{m^i,n^i}~\Gamma_{(2,2)}[^{n^i}_{m^i}]~=
{}~\Gamma_{(2,2)}[T,U].
\label{b22}\ee
When $h_i~\ne~0$ $(\gamma,~\delta)=(2h_in^i,~2h_im^i)$,  then the
$N=4$ supersymmetry is spontaneously broken to $N=2$ and the gauge
group is reduced to $U(1)^2 \times E_7 \times E_8$, as in orbifold
models.

The important difference between the $N=2$ model described above and
the orbifold models \cite{orbif} of order {\bf N} is in the
parameters $\gamma$ and $\delta$, which appear as arguments in
$\th$-functions. In the model in which some of the $N=4$ the
supersymmetries are broken spontaneously, $\gamma=2h_in^i$ and
$\delta=2h_im^i$ are not independent but are given in terms of the
$h_i$ and in terms of the charges $n^i,~m^i$ of the $\Gamma
(2,2)[^{n^i}_{m^i}]$ lattice.
In the standard symmetric  orbifolds of order {\bf N}, the arguments
$\gamma$ and $\delta$ ($\gamma=2l$/{\bf N} and  $\delta=2k$/{\bf
N} with  $l,k$ = 0,1,$\cdots$,{\bf N}-1) are independent  arguments;
their summation gives rise to the orbifold projections and to some
additional states in the twisted sector:

$$
\Z^{N=2}_{\rm orb}={1\over \tau_2|\eta|^4}\frac{1}{\bf N}
\sum_{\gamma,\delta}\sum_{\alpha,~\beta}~(-)^{\alpha+\beta +\alpha
\beta}\frac{\th^{2}[^{\alpha}_{\beta}]}{\eta^2}
\frac{\th[^{\alpha+\gamma}_{\beta+\delta}]}{\eta}
\frac{\th[^{\alpha-\gamma}_{\beta-\delta}]}{\eta}
{\Gamma_{(2,2)}\over |\eta|^4}~Z_{(4,4)}[^{\gamma}_{\delta}]
$$
\be
\times\frac{1}{2}\sum_{\bar \alpha, \bar \beta}
\frac{{\bar \th}^{6}[^{\bar \alpha}_{\bar \beta}]}{{\bar
\eta}^6}\frac{{\bar \th}[^{{\bar \alpha}+{\gamma}}_{{\bar
\beta}+{\delta}}]}{{\bar \eta}}\frac{{\bar \th}[^{{\bar \alpha}-{
\gamma}}_{{\bar
\beta}-{\delta}}]}{{\bar \eta}}\frac{1}{2}\sum_{\epsilon,
{}~\zeta}~\frac{{\bar
\th}^{8}[^{\epsilon}_{\zeta}]}{{\bar \eta}^8}.
\label{b23}\ee

In the language of orbifolds, the spontaneously broken theory,
$Z^{4\rightarrow 2}$, corresponds to a {\em freely acting orbifold}.
The possible (left--right symmetric) rotations we can use are of the
$Z_{\bf N}$ type with ${\bf N}=2,3,4,6$.
The model described above corresponds to ${\bf N}=4$.
The quantization condition becomes
\be
{\rm \bf N}~h_i~=~{\rm integer}
\label{b24}\ee
The $mod~2$ periodicity properties of the $\th$-functions in the
arguments,
\be
\th[^{a+2k}_{b+2l}]~=~\th[^{a}_{b}]~e^{i\pi la},
\label{b25}\ee
give us the possibility to write $Z^{4\rightarrow 2}$ in terms of the
orbifold language.
First we redefine the lattice charges $n^i={\rm \bf N}{\hat
n}^i+\gamma^i $ and $m^i={\rm \bf N}{\hat m}^i+\delta^i$. This
redefinition makes the arguments of the $\th$-functions independent of
${\hat n}^i$ and ${\hat m}^i$; they depend only on ${\hat
\gamma}~=~2h_i\gamma^i$ and ${\hat \delta}~=~2h_i\delta^i$.  We can
know perform a Poisson resummation on ${\hat m}^i$ to obtain the
following expression
for $\Z^{4\rightarrow 2}$
$$
\Z^{4\rightarrow 2}(F)= {1\over \tau_2|\eta|^4}{1\over {\bf
N}}\sum_{\gamma^i~\delta^i}~{1\over 2}~\sum_{\alpha,
\beta}~(-)^{\alpha+\beta
+\alpha\beta}~\frac{\th^{2}[^{\alpha}_{\beta}]}{\eta^2}
\frac{\th[^{\alpha+{\hat \gamma}}_{\beta+{\hat \delta}}]}{\eta}
\frac{\th[^{\alpha-{\hat \gamma}}_{\beta-{\hat \delta}}]}{\eta}
Z_{(4,4)}[^{\hat \gamma}_{\hat
\delta}]~\frac{\Gamma_{(2,2)}[^{\gamma^i}_{\delta^i}]}{|\eta|^4}
$$
\be
\times\sum_{{\bar \alpha}, {\bar \beta}}~
\frac{1}{2}~\sum_{\bar \alpha, \bar \beta}
\frac{{\bar \th}^{6}[^{\bar \alpha}_{\bar \beta}]}{{\bar \eta}^6}
{}~\frac{{\bar \th}[^{{\bar \alpha}+{\hat \gamma}}_{{\bar \beta}+{\hat
\delta}}]}{{\bar \eta}}~\frac{{\bar \th}[^{{\bar \alpha}-{\hat
\gamma}}_{{\bar \beta}-{\hat \delta}}]}{{\bar \eta}}
{}~\sum_{\epsilon, \zeta}\frac{1}{2}~\frac{{\bar
\th}^{8}[^{\epsilon}_{\zeta}]}{{\bar \eta}^8}
\label{b26}\ee
where
\be
\Gamma_{(2,2)}[^{\gamma^i}_{\delta^i}]=\sum\exp[i\pi{2\delta^i{\hat
m}_i \over {\bf {\bf N}}}+i\pi\tau~{1\over 2}
P^L_ig^{ij}P^L_j-i\pi{\bar \tau}{1\over 2}P^L_ig^{ij}P^L_j]
\label{b27}\ee
and
\be
P^L_i={\hat m}_i+({\hat n}^j+{\gamma^j \over {\bf N}})G_{ij}~~~~{\rm
and}~~~~
 P^R_i={\hat m}_i-({\hat n}^j+{\gamma^j \over {\bf N}})G_{ij}
\label{b28}\ee

The connection of $\Z^{4\rightarrow 2}$ with the freely acting
orbifolds gives us the way to switch all the  moduli of the $Z_{(4,4)}$
and thus move out of the extended symmetry of the $SO(8)$. This
extension can be done be replacing
$Z_{(4,4)}[^{\gamma}_{\delta}]$ which was defined at the fermionic
point by
\be
Z_{(4,4)}[^{0}_{0}]={\Gamma_{4,4}[T_{IJ}] \over
|\eta(\tau)|^8},~~~~{\rm}
\label{b29}\ee
while for $(\gamma,\delta)~\ne~(0,0)$
\be
Z_{(4,4)}[^{\gamma}_{\delta}]~=~Z_{(4,4)}^{twist}~
[^{\gamma}_{\delta}]~=~
\frac{1}{2}~\sum_{a,b}~\left|
{}~\frac{\th[^{a}_{b}]}{\eta}
{}~\frac{\th[^{a+2\gamma}_{b+2\delta}]}{\eta}
{}~\frac{\th[^{a+\gamma}_{b+\delta}]}{\eta}
{}~\frac{\th[^{a-\gamma}_{b-\delta}]}{\eta}\right|^2
\label{b30}\ee
The $(\gamma,\delta)~\ne~(0,0)$ part is the same at any point of the
moduli $T_{IJ}$ and is equal to the twisted part of the corresponding
orbifold partition function,
\be
Z_{(4,4)}^{twist}~[^{2k/\bf N}_{2k'/\bf
N}]~=~16\sin^4\left[{\pi\Lambda(k,k')\over \bf N}\right]
\left|{\eta \over \th\left[^{1+2k/\bf N}_{1+2k'/\bf
N}\right]}\right|^4,
\label{b31}\ee
where $\Lambda(k,k')=\Lambda(k',k)$ is :

$\bullet$ 1 for all non-trivial sectors for ${\bf N}=2$

$\bullet$ 1 for all nontrivial sectors for ${\bf N}=3$

$\bullet$ 2 for the sectors $(k,k')=(0,2),(2,0),(2,2)$ and 1 for
the rest for ${\bf N}=4$.

$\bullet$ 3 for the sectors $(0,3),(3,0),(3,3)$, 2 for
$(0,2),(2,0)$,
$(0,4),(4,0)$,$(2,2),(2,4)$, $(4,2),(4,4)$ and 1 otherwise for
${\bf N}=6$.

The models described above are special cases of a general class of
models
having the interpretation of freely acting orbifods of the N=4
heterotic string theory.
They are obtained in the following way.
Consider $\Gamma(6,22)$ as in (\ref{bb1}) and set the appropriate
moduli to
special values so that it factorizes as
\be
\Gamma_{6,22}\to \Gamma_{2,18}\;\Gamma_{4,4}
\label{b32}\ee
Now consider the orbifold that acts as a $Z_{\bf N}$ rotation on
$\Gamma_{4,4}$
and a translation by an ${\bf N}$-th lattice vector $\varepsilon/{\bf
N}$ with  $\varepsilon=(\vec\varepsilon_L;\vec\varepsilon_R,\vec
\zeta)$, on $\Gamma_{2,18}$.
$\vec\varepsilon_{L,R}$ are two-dimensional vectors while $\vec\zeta$
is a sixteen-dimensional vector.
The twisted blocks of the (4,4) piece are given by
$\Gamma_{4,4}[^0_0]=\Gamma_{4,4}$ and
\be
\Gamma_{4,4}[^h_g]=16\sin^4\left[{\pi\Lambda(h,g)\over \bf
N}\right]\left|{\eta^{2}\over
\th\left[^{1+2h}_{1+2g}\right]\th\left[^{1-2h}_{1-2g}\right]}
\right|^2\;\;\;{\rm for}\;\;\;(h,g)\not= (0,0)
\label{b33}\ee
with $h,g=1/{\bf N},\cdots,({\bf N}-1)/{\bf N}$.
Similarly for the (2,18) piece we obtain
\be
\Gamma^{\varepsilon}_{2,18}[^0_0]=\Gamma_{2,18}
\label{b34}\ee
\be
\Gamma^{\varepsilon}_{2,18}[^h_g]=\sum_{a\in {\cal
E}_{2,18}+h\varepsilon}e^{2\pi ia\cdot\varepsilon\over N}q^{{1\over 4}
a^T(M+L)a}\bar q^{{1\over 4}a^T(M-L)a}
\;\;\;{\rm for}\;\;\;(h,g)\not=(0,0)
\label{b35}\ee
Due to the accompanying translation, this is a freely acting orbifold.

The partition function can thus be written as
\be
Z^{4\to 2}_{\epsilon}={1\over \tau_2|\eta|^4}{1\over 2{\bf
N}}~\sum_{\alpha, \beta,h,g}~(-)^{\alpha+\beta +\alpha
\beta}~\frac{\th^{2}[^{\alpha}_{\beta}]}{\eta^2}
\frac{\th[^{\alpha+2h}_{\beta+2g}]}{\eta}
\frac{\th[^{\alpha-2h}_{\beta-2g}]}{\eta}
{\Gamma^{\varepsilon}_{2,18}[^h_g]
\Gamma_{4,4}[^h_g]\over \eta^6\bar\eta^{22}}
\label{b36}\ee
Modular invariance constrains the norm of $\varepsilon$:
\be
\varepsilon^2\equiv
2{\vec\e}_L\cdot\vec\varepsilon_R-\vec\zeta\cdot\vec\zeta\in 2Z
\ee
{}From the modular properties
\be
\tau\to\tau+1\;\;,\;\;\Gamma_{4,4}[^h_g]\to\Gamma_{4,4}[^h_{h+g}]
\;\;\;,\;\;\;\Gamma_{2,18}^{\varepsilon}[^h_g]\to e^{i\pi
h^2\varepsilon^2}~\Gamma^{\varepsilon}\_{2,18}[^h_{h+g}]
\label{b37}\ee
\be
\tau\to-{1\over \tau}\;\;,\;\;\Gamma_{4,4}[^h_g]\to\Gamma_{4,4}[^g_{h}]
\;\;\;,\;\;\;\Gamma_{2,18}^{\varepsilon}[^h_g]\to e^{-2i\pi
gh\varepsilon^2}~\Gamma^{\varepsilon}_{2,18}[^g_{-h}]
\label{b38}\ee
and
\be
\Gamma_{2,18}^{\varepsilon}[^{h+1}_g]=e^{-2i\pi g
\varepsilon^2}~\Gamma_{2,18}^{\varepsilon}[^h_g]\;\;\;,\;\;\;
\Gamma_{2,18}^{\varepsilon}[^{h}_{g+1}]=
\Gamma_{2,18}^{\varepsilon}[^{h}_g]
\ee
we obtain that
\be
{\varepsilon^2\over 2}=1\;\;{\rm mod}~{\bf N}^2
\label{b39}\ee
Moreover different lattice shifts do not always give different models
since
\be
\Gamma_{2,18}^{\varepsilon+{\bf N}\varepsilon'}[^{h}_g]=e^{-2\pi i{\bf
N}gh~\varepsilon\cdot\varepsilon'}~\Gamma_{2,18}^{\varepsilon}[^{h}_g]
\ee

The two types of constructions we have presented have complementary
features.
In the first approach, i.e. using a specific generalized boost at the
fermionic point, it is evident that there is a one-to-one
correspondence of states
between the original N=4 supersymmetric theory and the final
spontaneously
broken $N=4\to N=2$ theory. This is what should be expected during
spontaneous
symmetry breaking.
In the second, freely acting orbifold approach, we have a clear
geometrical
intuition about the spontaneously broken theory, which will be very
useful
to identify the type II dual.

We should remark here on the fate of T-duality.
Factorization of the (6,22) lattice gives a (2,18) lattice associated
with the
vector multiplets, with original $SO(2,18,Z)$ invariance.
There is also of (4,4) lattice associated with the neutral
hypermultiplets.
Its geometry, SO(4,4)/SO(4)$\times$SO(4) as well as O(4,4,Z) are
exact, since no perturbative or non-perturbative
corrections can modify them.
However, the discrete O(2,18,Z) symmetry is already broken by the
lattice shift $\e$.
Let $\Omega$ be an O(2,18,Z) matrix. Then the shifted lattice sum
transforms
as
\be
\Gamma_{2,18}^{\e}[^h_g](T_i)\to
\Gamma_{2,18}^{\Omega\e}[^h_g](T^{\Omega}_{i})
\ee
where $T^{\Omega}_i$ are the standard transformed (2,18) moduli.
The duality symmetry for a given groundstate (given $\e$) is the
O(2,18,Z)
subgroup that leaves $\e$ invariant up to even shifts on the lattice.
Broken transformations move us in the space of $\e$.
However, $\e^2$ is O(2,18,Z) invariant, and models with different
$\e^2$ cannot be related by ``broken" O(2,18,Z) transformations.

We can also give here the general mass formula for the massive
gravitini.
Inspection of the standard N=4 graviti vertex operators shows that two
of them are invariant while the other two transform, one with a phase
$e^{2\pi i/{\bf N}}$ and the other with $e^{-2\pi i/{\bf N}}$.
In order for them to survive in the spectrum they have to pair up with
a state
of the (2,18) lattice carrying momentum $p=(\vec m;\vec n,\vec Q)$ but
no oscillators (these will shift the mass to the Planck scale).
Since such a lattice state picks up a phase $e^{2\pi i \e\cdot p/{\bf
N}}$
one of the two massive gravitini will have momentum $p_1$ with the
property
that $p_1\cdot \e=1~~$mod$~{\bf N}$ while the other $p_2$ with
$p_2\cdot \e=-1$~~mod$~{\bf N}$.
The mass formulae given in (\ref {b18}) are special cases of the above.

Thus the mass of the gravitini are given by the holomorphic (2,18) mass
formula
(\ref{am36}). However there are several lattice vectors that satisfy
the above constraints. The ones that have the smallest mass are the
gravitini whereas the rest are Kaluza-Klein states of the usual or
massive
gravitini.
However, it is true that the statement of lowest mass depends on where
we are in the moduli space.
In fact there are explicit examples of models where supersymmetry is
restored
in different boundaries of moduli space, and it can be checked that the
gravitini that become massless there are different for the two
boundaries.

There is a related model with spontaneously broken N=4$\to$N=2
supersymmetry and a much smaller vector moduli space \cite{vw}.
It can be obtained by accompanying the $Z_2$ freely acting orbifold
projection described above with an operation that changes the sign of
the $E_8\times E_8$ lattice.
This gives a model with three vector multiplets (plus the graviphoton).

\vskip .5cm

There is an essential difference between the models with spontaneous
breaking of the $N=4\to N=2$ and the standard  $N=2$ orbifold models.

$\bullet$ First, in the spontaneously broken case, one expects an
effective
restoration of the $N=4$ supersymmetry in a corner of the moduli
space $T,~U$, where the two massive gravitinos become light,
$m_{3/2}\to 0$.

$\bullet$ Second, in the standard obifolds there is no restoration of
the $N=4$
supersymmetry at any point of the moduli space.

 If there is an effective restoration of the $N=4$ supersymmetry in
the spontaneously broken case, then one must find zero higher-genus
corrections to the coupling constants of the theory in the $N=4$
restoration limit $m_{3/2}\to 0$. This restoration phenomenon has
been checked in ref. \cite{kkpr1} where the  one-loop
corrections of the coupling constants were performed  for a class of
$Z_2$ models based on $E_8\times E_7\times SU(2)\times U(1)^2$ gauge
group. A more detailed discussion of the general heterotic models and
their type II duals will appear in ref. \cite{kkpr3}.
Here I will restrict myself to the case of $Z_2$ freely acting
orbifolds with $F~=h_1+Uh_2~=~1/2$. The  $m_{3/2}\to 0$ limit in this
class of models  corresponds  to the  corner of the moduli space
${\rm Im}T~{\rm Im}U \to \infty$, which implies an effective
decompactification of one  of the two coordinates  of
$\Gamma_{2,2}(T,U)$,  ($R_1\to \infty$ and $R_2$ arbitrary; ${\rm
Im}T \sim R_1 R_2,~~{\rm Im}U \sim R_1/R_2$). In this limit, $T,~U\to
\infty$, one expects vanishing corrections to the coupling constants
due to the effective  $N=4$  restoration. Using the explicit results
of ref. \cite{kkpr1},
\be
\Delta^{\rm free}_{(8,7)}={16 \pi^2 \over g^2_{E_8}}-{16 \pi^2 \over
g^2_{E_7}}=\delta\tilde b~{\rm log}\left[{\mu^2\over M_s^2}{\rm Im} T
{\rm Im}
U|\th_4(T)~\th_4(U)|^4 \right]+\label{b42}\ee
$$+\left({\delta b\over 3}-\delta\tilde b\right)\delta[T,U]
$$
where, $\tilde b_i$ are the massless $\beta$-functions of this model,
$b_i$ are
the massless $\beta$-functions of the standard $Z_2$ orbifold and
\be
\delta[T,U]=\int_{\cal F}{d^2\tau\over
\tau_2}\sum_{h,g}'\Gamma_{2,2}[^h_g]
\bar \sigma[^h_g]
\label{b43}\ee
with
\be
\bar\sigma[^h_g]={e^{2\pi i(g+h)}\over
16}\;{\bar\th^{12}[^{1+2h}_{1+2g}]\over
\bar\eta^{12}}\;\;,\;\;\sum_{h,g}'\bar \sigma[^h_g]=3
\label{b44}\ee
When $T$ and $U$  are large,
${\rm Im}T~{\rm Im}U~\gg~1$, due to the asymptotic behaviour of
$\th_4(T)=1+{\rm {\cal O}}(e^{-{i\pi~T}})$:
\be
\Delta^{\rm free}_{(8,7)}~~\rightarrow ~~\delta \tilde b~{\rm
log}~({\rm Im} T~{\rm Im} U)+{\cal O}(1/{\rm Im} T{\rm Im}U).
\ee
The logarithmic contribution is an  artefact due to the infrared
divergences.
In fact by turning on Wilson lines appropriately (e.g. small Higgs
vev's of the vector multiplets), we can arrange that there are no
charged states with masses $\mu^2_W ~\sim ~|W|^2/{\rm Im} T~{\rm Im}
U~$ below $m^2_{3/2}$. In this case the logarithmic term becomes:
\be
\delta b~{\rm log}~(\mu^2~{\rm Im}T~{\rm Im} U) ~~\rightarrow
{}~~\delta
b~{\rm log}~{\mu^2_W \over m^2_{3/2}+\mu^2_W  }
\sim~{\rm \cal O}
\left({m^2_{3/2} \over \mu^2_W } \right);
\ee
the logarithmic divergence thus disappears and the thresholds vanish,
which shows the restoration of the $N=4$ supersymmetry in the light
massive gravitino limit as expected.
 In the calculation of individual couplings, there is an extra
contribution $Y(T,U)$, which is ``universal" for $g_{E_8}$ and
$g_{E_7}$; the explicit calculation in
\cite{kkpr1}, \cite{kkpr3} shows that $Y(T,U)$ behaves  like
\be
Y(T,U ) \rightarrow {m^2_{3/2}\over M^2_s}\;\;\;{\rm as}\;\;\;
m_{3/2}\to 0.
\label{b48}\ee
Thus individual couplings also vanish in the limit $m_{3/2}\to 0$.

 In the standard orbifold with $N=2$ space-time supersymmetry, the
corrections to the coupling constants have a different behaviour for
$T,~U\gg 1$ \cite{dkl}:
\be
\Delta^{orb}_{(8,7)}=\delta b~\log\left[\mu^2~{\rm Im}T{\rm Im}
U|\eta(T)\eta(U)|^4 \right].
\label{b40}\ee
When $T,U$ is large, ${\rm Im}T {\rm Im}U \gg~1$, \cite{ns}
\be
\Delta^{orb}_{(8,7)}~~\rightarrow ~~\delta b~\left[{\pi \over 3}({\rm
Im}T+{\rm Im}U)~+~ \log~{|W|^2 \over M^2_s} \right]
+{\rm finite~terms}.
\label{b41}\ee
while the universal piece blows up linearly with the volume
\cite{kkpr2,ns}.

So, in the standard orbifolds, the correction to the coupling
constants grows linearly with the  five-dimensional volume. This
shows that the $N=2$ supersymmetry is ``not extended" in the
decompactification limit $R_1\to\infty$. On the other hand there is
an extension of the supersymmetry in the freely acting orbifold case.

In the opposite limit ${\rm Im}T {\rm Im}U\to 0$, the situation is
different:

 i) In  the freely acting orbifold the two massive gravitinos
become superheavy: $m_{3/2}\to \infty$ in the limit ${\rm Im}T {\rm
Im}U\to 0$.

ii) In the  standard orbifold, thanks to the duality symmetry
$R_i\to 1/R_i$ the behaviour $T,U\to 0$ is identical to the dual
model with  $T'=-1/T,~U'=-1/U~\to \infty$ and thus
\be
\Delta^{orb}_{(8,7)}(T,U,W)=\Delta^{orb}_{(8,7)}(T',U',W')\rightarrow
\delta b\left[{\pi \over 3}({\rm Im}T'+{\rm Im}U')+\log{|W'|^2 \over
M^2_s}\right]+{\rm finite}
\label{b41o}\ee

In the freely acting orbifolds, the $SO(2,2;Z)$ duality symmetry  is
reduced to a smaller subgroup due to the $Z_2$ action on the lattice.
Thus one expects non-restoration of the $N=4$ supersymmetry  in this
limit ( $T,~U\to 0$ $m_{3/2}\to\infty$).
Using
\be
{\rm Im}T|\th_4(T)|^4={\rm Im}T'|\th_2(T')|^4, ~~~T'=-{1 \over
T}.
\label{b41th}\ee
we obtain in the limit
\be
\Delta^{free}_{(8,7)}\rightarrow \delta b\left[{\pi \over
3}({\rm Im}~T'+{\rm Im}~U')+ \log{|W'|^2 \over M^2_s}\right]
+{\rm finite~terms}.
\label{b41as}\ee
Note that $\delta\tilde b$ has disappeared.

It is interesting to observe that the $m_{3/2}\to\infty$ limit
\cite{kkpr1} of the freely acting orbifolds corresponds  to a corner
in the moduli space of $T,U$  where the two classes of theories (the
freely  and non-freely acting orbifolds)  ``touch" each other.
Both theories are effectively five-dimensional. Thus the
five-dimensional standard $N=2$ orbifolds can be viewed as an
$m_{3/2}\to \infty$ limit of some spontaneously broken $N=4$ models.

\section{BPS States in $N=4\to N=2$ ground states}
\setcounter{equation}{0}

\def\l{\lambda}
Let us consider the interesting question concerning the BPS spectrum of
the theories where $N=4$ is spontaneously broken to N=2.
In the original heterotic N=4 theory, there are only short multiplets
in the perturbative spectrum.
Their multiplicities can be easily counted using helicity supertrace
formulae \cite{bk}.
In particular, the supertrace of helicity to the power four counts
exactly
the multiplicities\footnote{More accurately this is an ``index",
namely, the difference between integer spin minus  half-integer spin
multiplicities.}
of N=4 short (massless or massive) multiplets.
Introduce the helicity generating partition function
\be
Z^{het}_{N=4}(v,\bar v)=Str[q^{L_0}\;\bar q^{\bar L_0}e^{2\pi
iv\l_R-2\pi i\bar v\l_L}]={1\over
2}\sum_{\a\b}(-1)^{\a+\b+\a\b}{\vartheta[^{\a}_{\b}](v)
\vartheta^3[^{\a}_{\b}]\over\eta^{12}\bar
\eta^{24}}\xi(v)\bar\xi(\bar v){\Gamma_{6,22}\over {\rm
Im}\tau}=\label{cc18}\ee
$$={\vartheta^4_1(v/2)\over
\eta^{12}\bar \eta^{24}}\xi(v)\bar\xi(\bar v){\Gamma_{6,22}\over {\rm
Im}\tau}
$$
The physical helicity in closed string theory $\l$ is a sum of the
left helicity $\l_L$ and the right helicity $\l_R$
\be
\xi(v)=\prod_{n=1}^{\infty}{(1-q^n)^2\over (1-q^ne^{2\pi iv})
(1-q^ne^{-2\pi iv})}={\sin\pi v\over \pi}{\vartheta_1'\over
\vartheta_1(v)}\;\;\;\,\;\;\;\xi(v)=\xi(-v)
\label{cc3}\ee
counts the contributions to the helicity due to the world-sheet
bosons.
If we define
\be
Q={1\over 2\pi i}{\partial\over \partial v}\;\;\;,\;\;\;\bar
Q=-{1\over 2\pi i}{\partial\over \partial \bar v}
\label{cc4}\ee
then
\be
B_4=\langle \lambda^4\rangle =(Q+\bar Q)^4 Z^{het}_{N=4}(v,\bar
v)|_{v=\bar v=0}={3\over 2}{\Gamma_{6,22}\over \bar\eta^{24}}
\label{bb39}\ee
The numerator provides the mass formula while the denominator
$1/\bar\eta^{24}$ provides the multiplicities.
More precisely define
\be
{1\over \eta^{24}}={1\over q}+\sum_{n=1}^{\infty}d(n)q^{n}={1\over
q}+24 +324 q+{\cal O}(q^2)
\label{bb40}\ee
Then at the mass level, $M^2={1\over 4}p_L^2$, with
\be
{1\over 2}q_e^2\equiv
\vec m\cdot\vec n-\vec Q\cdot \vec Q/2
\label{bb41}\ee
the multiplicity is $d(q_e^2)$.
This multiplicity formula was generalized to dyonic states \cite{dvv}
which are characterized, apart from their electric charges, also by
magnetic charges $\vec m',\vec n',\vec Q'$.
Define
\be
{1\over 2} q_m^2=\equiv\vec m'\cdot\vec n'-\vec Q'\cdot \vec Q'/2
\label{bb42}\ee
\be
q_e\cdot q_m\equiv \vec m\cdot\vec n'+\vec m'\cdot\vec m-
\vec Q\cdot \vec Q'
\label{bb43}\ee
and consider the genus-2 $\eta$ function
\be
\eta\left[\matrix{T&V\cr V&U\cr}\right]^{-24}=
\sum_{k_1,k_2,k_3}d(k_1,k_2,k_3)e^{2\pi i(k_1 T+k_2 U+k_3 V)}
\ee
Then, the non-perturbative multiplicities are given by
$d(q_e^2/2,q_m^2/2,q_e\cdot q_m)$.
In general we expect $B_4$ for heterotic N=4 groundstates to transform
as
\be
\tau\to\tau+1\;\;:\;\;B_4\to B_4\;\;\;,\;\;\;\tau\to -{1\over
\tau}\;\;:\;\;
B_4\to \tau^4 ~B_4
\label{bbb443}\ee

Consider now the spontaneously broken $N=4\to N=2$ theories, and for
simplicity
we will restrict to the $Z_2$ case.
Massless multiplets $M_0^{\lambda}$ have the following helicity content
\be
\pm\left(\lambda\pm{1\over 2}\right)+2(\pm \lambda)
\label{bbb43}\ee
$M_0^0$ is the hypermultiplet, $M_0^{1/2}$ is the vector multiplet
while $M_0^{3/2}$ is the supergravity multiplet.
The massive BPS multiplets have the following $SO(3)$ spin content
\be
M^j\;\;:\;\;[j]\otimes ([1/2]+2[0])
\label{bbb44}\ee
and contain $2(2j+1)$ bosonic and an equal number of fermionic states.
Finaly the generic long massive multiplet has the following SO(3)
content
\be
L^j\;\;:\;\;[j]\otimes ([1]+4[1/2]+5[0])
\label{bbb45}\ee

In the N=2 case, only the short (BPS) multiplets are picked up by the
supertrace of helicity squared, $B_2=\langle\lambda^2\rangle$. We have
\be
B_2(M_0^{\lambda})=(-1)^{2\lambda}\;\;\;,\;\;\;
B_{2}(M^j)=(-1)^{2j+1}(2j+1)/2\;\;\;,\;\;\;B_2(L^j)=0
\label{bbb46}\ee

A direct computation along the lines of \cite{bk} gives
\be
\tau_2~B_2=\tau_2~\langle \l^2\rangle
=\Gamma_{2,18}[^0_1]{\bar\th^2_3\bar\th_4^2\over
\bar\eta^{24}}-\Gamma_{2,18}[^1_0]{\bar\th^2_2\bar\th_3^2\over
\bar\eta^{24}}-\Gamma_{2,18}[^1_1]{\bar\th^2_2\bar\th_4^2\over
\bar\eta^{24}}
\label{bb44}\ee
$$={\Gamma_{2,18}[^0_0]+\Gamma_{2,18}[^0_1]\over 2}\bar
F_1-{\Gamma_{2,18}[^0_0]-\Gamma_{2,18}[^0_1]\over 2}\bar
F_1-{\Gamma_{2,18}[^1_0]+\Gamma_{2,18}[^1_0]\over 2}\bar
F_+-{\Gamma_{2,18}[^1_0]-\Gamma_{2,18}[^1_0]\over 2}\bar F_-
$$
with
\be
\bar F_1={\bar\th^2_3\bar\th_4^2\over \bar\eta^{24}}\;\;\;,\;\;\;
\bar F_{\pm}={\bar\th^2_2(\bar\th_3^2\pm\bar\th_4^2)\over
\bar\eta^{24}}
\label{bb46}\ee
For all N=2 heterotic groundstates $B_2$ transforms as
\be
\tau\to\tau+1\;\;:\;\;B_2\to B_2\;\;\;,\;\;\;\tau\to -{1\over
\tau}\;\;:\;\;
B_4\to \tau^2 ~B_4
\label{bbb444}\ee

All functions $\bar F_i$ have positive coefficients and have the
generic
expansions
\be
F_{1}={1\over q}+\sum_{n=0}^{\infty}d_{1}(n)q^n={1\over q}+16+156
q+{\cal O}(q^2)
\label{bb47}\ee
\be
F_{+}={8\over q^{3/4}}+q^{1/4}\sum_{n=0}^{\infty}d_{+}(n)q^n={8\over
q^{3/4}}+8q^{1/4}(30+481 q+{\cal O}(q^2))
\label{bb48}\ee
\be
F_{-}={32\over q^{1/4}}+q^{3/4}\sum_{n=0}^{\infty}d_{-}(n)q^n=
{32\over q^{1/4}}+32q^{3/4}(26+375q+{\cal O}(q^2))
\label{bb49}\ee
Also the lattice sums $(\Gamma_{2,18}[^h_0]\pm\Gamma_{2,18}[^h_1])/2$
have
positive multiplicities.
Overall plus signs correspond to vector-like multiplets, minus
signs to hyper-like multiplets.
The contribution of the generic massless multiplets is given by the
constant coefficient of $F_1$ and agrees with what was expected:
16=20-4
since we have the supergravity multiplet and 19 vector multiplets
contributing 20 and 4 hypermultiplets contributing --4.
Turning off all the Wilson lines and restoring the $E_7\times E_8$
group, the above result becomes
\be
\langle \l^2\rangle
=\Gamma_{2,2}[^0_1]{\bar\th^4_3\bar\th_4^4(\bar\th_3^4+\bar\th_4^4)
\bar E_4\over 2\bar\eta^{24}}-\Gamma_{2,2}[^1_0]
{\bar\th^4_2\bar\th_3^4(\bar\th_2^4+\bar\th_3^4)\bar E_4\over
2\bar\eta^{24}}-\Gamma_{2,2}[^1_1]{\bar\th^4_2\bar\th_4^4
(\bar\th_2^4-\bar\th_4^4)\bar E_4\over 2\bar\eta^{24}}
\label{bb45}\ee

Let us analyse the BPS mass formulae associated with (\ref{b29}) with
${\bf N}=2$.
We will use the notation of the previous chapter with the (2,18) shift
vector
$\e=(\vec\e_L;\vec\e_R,\vec\zeta)$ satisfying the constraint
(\ref{b39}).
Using the results of section 2 we can write the mass formulae
associated
to the lattice sums above.
For $h=0$
the mass formula is
\be
M^2={|-m_1U+m_2+Tn_1+(TU-{1\over 2}
\vec W^2)n_2+\vec W\cdot \vec Q|^2\over 4 \;S_2
\left(T_2U_2-{1\over 2}{\rm
Im}\vec W^2\right)}
\label{bb50}\ee
where $\vec W$ is the sixteen-dimensional complex vector of Wilson
lines.
When the integer
\be
\rho=\vec m\cdot \vec \e_R+\vec n\cdot\e_L-\vec Q\cdot\vec\zeta
\label{bb51}\ee
is even these states are vector multiplet-like with multiplicity
function
$d_1(s)$ of (\ref{bb47})
and
\be
s=\vec m\cdot\vec n-{1\over 2}\vec Q\cdot\vec Q
\label{bb52}\ee
and when $\rho$ is odd these states are hypermultiplet-like with
multiplicities $d_1(s)$.
In the $h=1$ sector the mass formula is
\be
M^2={|-(m_1+{\e_L^1\over 2})U+(m_2+{\e_L^2\over 2})+T(n_1+{\e_R^1\over
2})+(TU-{1\over 2}
\vec W^2)(n_2+{\e_R^2\over 2})+\vec W\cdot (\vec Q+{\vec \zeta\over
2})|^2\over 4 \;S_2
\left(T_2U_2-{1\over 2}{\rm
Im}\vec W^2\right)}
\label{bb53}\ee
The states with $\rho$ even are hypermultiplet-like with multiplicities
$d_+(s')$ with
\be
s'=\left(\vec m+{\vec\e_L\over 2}\right) \cdot\left(\vec n
+{\vec \e_R\over 2}\right)-{1\over 2}\left(\vec Q+{\vec\zeta\over
2}\right)\cdot\left(\vec Q+{\vec\zeta\over 2}\right)
\label{bb54}\ee
while the states with $\rho$ odd are hypermultiplet like with
multiplicities
$d_-(s')$.

Let us here consider symmetry enhancement in the presence of shift
vectors.
For simplicity we will set the Wilson lines to zero and ignore the
charged sector. The case with non-zero Wilson lines is straightforward
but more involved.
Let us fist consider the untwisted sector (h=0).
According to the above analysis the masses are given by the unshifted
mass formula
(\ref{bb50}) and they are vector multiplets when  $\rho$ is even and
hypermultiplets when $\rho$ is odd.
Now the points where the standard (2,2) mass vanishes are well known.
They are O(2,2,Z) images of $T=U$.
In the unshifted case, O(2,2,Z) is a symmetry and up to it there is
a single enhanced symmetry point, namely $T=U$.
When there are non-trivial shifts involved, the T-duality group is
smaller as we discussed in the previous section,
and there is a finite little group $G^{\varepsilon}$ which acts
non-trivially
on the moduli.
Then the points where we obtain massless states are images of $T=U$
under $G^{\varepsilon}$.
We will pick
$T=U$ the other points can be obtained by duality.
There are two configurations with zero mass
given by $m_1=n_1=\pm 1$, all the rest being zero.
For both states $|\rho|=|\e_L^1+\e_R^1|$. Depending on it being even or
odd these states are either vector multiplets that enhance the gauge
group
$U(1)^2\to SU(2)\times U(1)$ or hypermultiplets charged under one of
the
$U(1)$'s.

Let us now look for states becoming massless in the twisted ($h=1$)
sector.
(\ref{bb46}) implies that massless states from the twisted sector are
always
hypermultiplets.
At $T=U$ a potentially massless state must have
$m_1+\e_L^1/2=n_1+\e_R^1/2$
and $\e_L^2,\e_R^2$ even integers.
If $\rho$ is even then for the states to be massless they must satisfy
$s'=3/4$.
{}From (\ref{bb48}) we deduce that in such a case  there will be 8
massless hypermultiplets.
If $\rho$ is odd, then the state will be massless if $s'=1/4$ and
(\ref{bb49})
implies that there will be 32 massless hypermultiplets.

\section{$N=4\rightarrow N=1$ Spontaneous Supersymmetry
 Breaking}
\setcounter{equation}{0}

Using the connection between the freely acting orbifolds and the
spontaneous breaking $N=4\rightarrow N=2$, we can proceed to further
break  the supersymmetry to $N=1$.
We will restrict ourselves  to the  case where the possible quantized
 parameters are of order  {\bf N}=2,  $2|h^i|=1$. In that case the
spontaneously broken $N=4\rightarrow N=1$ theory is related
to $Z^2\times Z^2 $ freely acting orbifolds; the $Z^2\times
Z^2$ acts simultaneously as a  rotation on the coordinates
$\Phi^I,~\Phi^J$ and $\Psi^I,~\Psi^J$ of the two complex planes and
as  a translation on the third complex plane $\Phi^L$. Denoting  by
$\Phi_A,~A=1,2,3$, the complex internal coordinates and by
$\chi_A,~A=1,2,3$, the three complex fermionic world-sheet
superpartners, the non-trivial actions of the orbifold are:
\vskip .2cm
1) $\Phi_1 \rightarrow \Phi_1+2\pi h_1$,$~~~(\Phi_2,~\chi_2)~
\rightarrow e^{~2 \pi
ih_1}(\Phi_2,~\chi_2)$,$~~~(\Phi_3,~\chi_3)~\rightarrow e^{-2\pi
ih_1}(\Phi_3,~\chi_3)$.
\vskip .2cm

2) $\Phi_2 \rightarrow \Phi_2+2\pi
h_2$,$~~~(\Phi_1,~\chi_1)~\rightarrow
e^{2\pi  ih_2} (\Phi_1,~\chi_1)$,$~~~(\Phi_3,~\chi_3)\rightarrow
e^{-2\pi  ih_2}(\Phi_3,~\chi_3)$.

\vskip .2cm
3) $\Phi_3 \rightarrow \Phi_3+2\pi
h_3$,$~~~(\Phi_1,~\chi_1)~\rightarrow
e^{2\pi  ih_3} (\Phi_1,~\chi_1)$,$~~~(\Phi_2,~\chi_2)~ \rightarrow
e^{-2\pi  ih_3}(\Phi_2,~\chi_2)$.
\vskip .2cm

In order to obtain the partition function and define the theory, we
need to introduce
the ``shifted" and ``twisted" characters of the three complex
coordinates.
We denote by $(\gamma_A,~\delta_A)$ the translation shifts and by
$(H_A,~G_A)$ the rotation twists.
These orbifold blocks are derived in Appendix A.

We are now in a position to construct $N=4\to N=1$ models.
We will display below the partition function of a  model with one
unbroken and three
spontaneously broken supersymmetries, $N=4\rightarrow N=1$ ( the
unbroken gauge group of this example is $E_8\times E_6 \times
U(1)^2$:

\be
\Z^{4\rightarrow 1}= {1\over \tau_2|\eta|^4}
\frac{1}{4}\sum_{h_i,g_i}
Z_1\left[^{h_1 ;h_2}_{g_1;g_2}\right]~
Z_2\left[^{h_2 ;h_1+h_2}_{g_2;g_1+g_2}\right]~
Z_3\left[^{h_1+h_2;h_1}_{g_1+g_2;g_1}\right]
{}~\frac{1}{2}\sum_{\epsilon,
\zeta}\frac{{\bar
\th}^{8}[^{\epsilon}_{\zeta}]}{{\bar \eta}^8}
\label{c5}\ee
$$
{1\over 2}\sum_{\alpha,
\beta}(-1)^{\alpha+\beta}\frac{\th[^{\alpha}_{\beta}]}{\eta}
\frac{\th[^{\alpha+h_2}_{\beta + g_2}]}{\eta}
\frac{\th[^{\alpha -h_1-h_2}_{\beta -g_1-g_2}]}{\eta}
\frac{\th[^{\alpha +h_1}_{\beta + g_1}]}{\eta}~
\frac{1}{2}\sum_{\bar \alpha, \bar \beta}
\frac{{\bar \th}^{5}[^{\bar \alpha}_{\bar \beta}]}{{\bar \eta}^5}
\frac{{\bar \th}[^{{\bar \alpha} + h_2}_{{\bar\beta} + g_2}]}
{{\bar \eta}}\frac{{\bar \th}[^{{\bar \alpha}-h_1-h_2}_{{\bar
\beta}-g_1-g_2}]}{{\bar \eta}}\frac{{\bar \th}[^{{\bar \alpha} + h_1}
_{{\bar\beta} + g_1}]}{{\bar \eta}}.
$$
The massless chiral multiplets (appart from the universal ones) are the
following (E$_8$,E$_6$,U(1),U(1)')

$\bullet$  One (1,27,1/2,1/2) + c.c.

$\bullet$  One (1,27,-1/2,1/2) + c.c.

$\bullet$  One (1,27,0,-1) + c.c.

$\bullet$  One (1,1,-1/2,3/2) + c.c.

$\bullet$  One (1,1,+1/2,3/2) + c.c.

$\bullet$  One (1,1,+1,0) + c.c.

The spectrum is non-chiral.

It is easy to see that the partition function $\Z^{4\rightarrow 1}$
can be decomposed in four sectors (we write $g_3=-(g_1+g_2)$,
$h_3=-(h_1+h_2)$):

$\bullet$ {\bf The $N=4$ sector}, with  no rotations  or translations
in
all three complex planes  ($(h_A,~g_A)=(0,0)$)

$\bullet$ {\bf Three  $N=2$ sectors}, with a non-zero translation in
one of the complex planes and  opposite non-zero rotations  in the
remaining two complex planes.

The contribution to the partition function of the $N=4$ sector is one
quarter of the $N=4$ partition function with  lattice momenta in the
reduced $\Gamma (2,2)^3$ lattice.
The contribution of the other three $N=2$ sectors are equal sector by
sector
to the corresponding $N=4 \rightarrow N=2$ partition function divided
by a factor of 2. The  untwisted complex plane lattice momenta
correspond to the shifted  $\Gamma_{(2,2)}~[^{\gamma_A}_{\delta_A}]$
lattice.
The moduli-dependent corrections to the gauge couplings can be easily
determined by combining the results of the individual $N=2$ sectors:
\be
{16 \pi^2 \over g^2_{E_8}}~-{16 \pi^2 \over
g^2_{E_6}}~=~\Delta_{(8,6)}={1\over 2}\sum^3_{A=1}~\Delta^A_{(8,7)},
\label{c7}\ee
where the expressions of the $\Delta^A_{(8,7)}$ are given in
(\ref{b33}).

 As we  mentioned in the $N=4\rightarrow N=2$ spontaneous breaking,
one expects
a restoration of the $N=4$ supersymmetry in the limit in which the
massive gravitini become massless; in order to prove the $N=4$
restoration in the $N=4\rightarrow N=1$ defined above  as a
$Z^2\times Z^2$ freely acting orbifold, we need to identify the three
massive gravitini and express their masses in terms of the moduli
fields and the  three  $R$-symmetry charges $q_i~(i=1,2,3)$:
\be
m^2_{3/2}(q_i)=\frac{|q_2-q_3|^2}{4{\rm Im}T_1{\rm Im}U_1}+
\frac{|q_3-q_1|^2}{4{\rm Im}T_2{\rm Im}U_2}+
+\frac{|q_1-q_2|^2}{4{\rm Im}T_3{\rm Im}U_3}
\label{c8}\ee
with $|q_0+q_1+q_2+q_3|=1~~{ \rm and}
{}~~|q_i|=|q_0|={1 \over 2}$
$q_0$ being the left-helicity charge. Using the
above expression, one finds the desired result:
$$
(m^2_{3/2})_1~=~\frac{1}{4~{\rm Im}~T_2~{\rm
Im}~U_2}~+~\frac{1}{4~{\rm Im}~T_3~{\rm Im}~U_3}, ~~~~~
$$
\be
(m^2_{3/2})_2~=~\frac{1}{4~{\rm Im}~T_3~{\rm
Im}~U_3}~+~\frac{1}{4~{\rm Im}~T_1~{\rm Im}~U_1}, ~~~~~
\label{c9}\ee
$$
(m^2_{3/2})_3~=~\frac{1}{4~{\rm Im}~T_1~{\rm
Im}~U_1}~+~\frac{1}{4~{\rm Im}~T_2~{\rm Im}~U_2}, ~~~~~
$$
and  $(m^2_{3/2})_0~=~0$.

The three massive gravitini become massless in the
decompactification limit
${\rm Im}~T_I~{\rm Im}~U_I $ $\to \infty$, $~I=1,2,3$, with ratios
${\rm Im}~T_I/{\rm Im}~U_I$ fixed. Thus the full restoration of the
$N=4$  effectively takes place in seven dimensions.
Partial restoration of an $N=2$ supersymmetry can happen in six
dimensions when ${\rm Im}T_I~{\rm Im}~U_I$ $\to\infty, ~~I=1,2$; in
this limit
$(m^2_{3/2})_0 ~=~ 0$ and  $(m^2_{3/2})_3 \rightarrow 0$.

\section{N=2$\rightarrow$ N=1  spontaneous SUSY breaking }
\setcounter{equation}{0}

Using similar techniques as before, it is possible to construct N=2
models  with one of the supersymmetries to be spontaneously broken,
N=2$\rightarrow$N=1. In this class of models the restoration of
N=2 takes place in six dimensions. No further restoration of
supersymmetry is possible. Examples
can be obtained as in $(T^2\otimes K_3)/ Z^2_{f}$ orbifold
compactification in which the $Z^2_{f}$ is freely acting (this is known
as the Enriques involution of $K_3$). Moreover, as we will see
chirality
can be present in the N=1 groundstate.
A representative example of this class of models is the one in which
the $K_3$ compactification is chosen to be at the  orbifold point
$T^4/Z^o_2\sim K_3$  (we denote by  $Z_2^o$ the orbifold group and by
$Z_2^f$ that which corresponds to the freely acting orbifold). We
will give below  the partition function that corresponds to
this construction. From the  explicit expression we can directly
verify  the effective restoration of $N=2$ supersymmetry in the
large-volume limit of $K_3$. Using the $Z_2^o \times Z_2^f$ orbifold
notation, the partition function of the $( T^2 \otimes
T^4/Z^o_2)/Z_2^{f}$ model is:

\be
\Z^{2\rightarrow 1}={1 \over \tau_2|\eta|^4}
\frac{1}{4}\sum_{h_f,g_f,h_o,g_o}
Z_1\left[^{0 ; h_f}_{0 ; g_f}\right]
Z_2\left[^{h_f  ;h_o}_{g_f ; g_o}\right]
Z_3\left[^{h_f ;-h_f-h_o}_{ g_f ; -g_f-g_o}\right]~\frac{1}{2}
\sum_{\epsilon, \zeta}~\frac{{\bar
\th}^{8}[^{\epsilon}_{\zeta}]}{{\bar \eta}^8}
\label{c10}\ee
$$
{1\over 2} \sum_{\alpha, \beta} (-)^{\alpha+\beta}
\frac{\th[^{\alpha}_{\beta}]}{\eta}\frac{\th[^{\alpha+
h_f}_{\beta + g_f}]}{\eta}
\frac{\th[^{\alpha + h_o}_{\beta + g_o}]}{\eta}
\frac{\th[^{\alpha -h_f-h_o}_{\beta -h_f-h_o}]}{\eta}
\frac{1}{2}~\sum_{\bar \alpha, \bar \beta}
\frac{{\bar \th}^{5}[^{\bar \alpha}_{\bar \beta}]}{{\bar \eta}^5}
\frac{{\bar \th}[^{{\bar \alpha} + h_f}_{{\bar
\beta} +g_f}]}{{\bar \eta}}
\frac{{\bar \th}[^{{\bar \alpha} + h_o}_{{\bar \beta} + g_o}]}{{\bar
\eta}}
\frac{{\bar \th}[^{{\bar \alpha} - h_f - h_o}_{{\bar
\beta}-g_f - g_o}]({\bar \tau})}{{\bar \eta}}
$$
In the above expression, the  parameters $(h_f,~g_f)$ and
$(h_o,~g_o)$    correspond to  $Z^2_f$ and  $Z^2_o$ respectively. The
unbroken gauge group of this model is the $E_8\otimes E_6 \otimes
U(1)^2$. Switching on continuous or discrete Wilson lines, we can
construct a large class of models with different gauge group but with
a universal behaviour
with respect to the $N=2$  restoration at the large moduli limit; the
 massive gravitino of the broken $N=2$ becomes massless when (${\rm
Im}~T_2~{\rm Im}~ U_2$ and ${\rm Im}~T_3~{\rm Im}~U_3~$ large).
\be
(m^2_{3/2})_1~=~\frac{1}{4~{\rm
Im}~T_2~{\rm Im}~U_2}~+~\frac{1}{4~{\rm Im}~T_3~{\rm
Im}~U_3}\;\;\;,\;\;\;
(m^2_{3/2})_0~=~0.
\label{c11}\ee

The massless spectrum coming from the untwisted sectors is non-chiral.
However, here we do obtain chiral fermions from the twisted sectors.
In particular we have 16 copies of the $27$ of $E_6$.
This implies that in string theory, unlike field theory, chirality can
appear
after spontaneous breaking of extended supersymmetry (N=2 in our
example above).
Moreover, we can vary the supersymmetry breaking scale without breaking
the gauge group with chiral representations ($E_6$ here).
It is a very interesting open problem if one can produce a chiral
spectrum in
the spontaneous breaking of $N=4\to N=1$.
We have no concrete example of this but also no counter-argument
either.

An easy way to view this groundstate is as an orbifold of the
original $N=4$ theory by the following non-trivial $Z_2\times Z_2$
elements: $(1,r,r), ~(r,rt,t), ~(r,t,rt)$, ($r$ stands for
``$\pi$-rotation" and $t$ for half-lattice translation);
$(1,r,r)$ has four fixed planes while the others have none.
Because of  the $N=2$ restoration phenomenon, we expect that the only
non-vanishing
corrections to the gauge coupling constants are those that correspond
to the
$N=2$ sector with $(h_f,~g_f)=(0,~0)$ and $(h_o,~g_o)\ne (0,~0)$.
Indeed in this sector the $Z^2_o$ acts trivially  on the
$\Gamma_{(2,2)}(T_1,~U_1)$ lattice as in the usual orbifolds. On the
other hand, in the  remaining two $N=2$ sectors,

i) $(h_o, g_0)= (0, 0)$,  $(h_f, g_f)\ne(0, 0)$

ii) $(h_o, g_0)+(h_f,g_f)= (0, 0)$, $(h_f, g_f)\ne(0, 0)$.

In both sectors the corresponding  $Z^2$ acts without fixed points
because of the simultaneous non-trivial shift $(h_f,~g_f)$ on the
corresponding $\Gamma_{(2,2)}(T_A,~U_A), ~~A=2,3$, lattice.

The moduli-dependent corrections to the gauge couplings can be easily
determined by combining the results of the individual $N=2$ sectors:
\be
\Delta_{(8,6)}={16 \pi^2 \over g^2_{E_8}}-{16 \pi^2 \over
g^2_{E_6}}=
{1\over 2} \left(~\Delta^1_{(8,7)} +\Delta^2_{(8,7)} +
\Delta^3_{(8,7)}\right) ,
\label{c12}\ee
where the $\Delta^A_{(8,7)}$ are the threshold corrections of the
three $N=2$ sectors:
\be
\Delta^1_{(8,7)}=(b^1_8-b^1_7)~{\rm log}\left[ |\mu|^2{\rm Im}
T_1{\rm Im} U_1 |\eta(T_1)\eta(U_1)|^4 \right]
\label{c13}\ee
$$
\rightarrow (b^1_8-b^1_7) \left[{\pi \over 3}(~{\rm
Im}T_1 + {\rm Im}U_1) + {\rm log} |\mu|^2 {\rm Im} T_1 {\rm Im}
U_1\right]
$$
which corresponds to the threshold corrections of the standard
orbifolds.

On the other hand $\Delta^A_{(8,7)}$ for $A=2,3$ will correspond to
the threshold corrections of freely acting orbifolds which have
different  behaviour in the large-moduli limit:
\be
\Delta^A_{(8,7)}=(b^A_8-b^A_7)~{\rm log}~\left[\mu^2{\rm
Im}T_A{\rm Im}U_A\right]
+(b^A_8-b^A_7){\rm log}\left[ |\th_4(T_A)\th_4(U_A)|^4 \right]~
\label{c14}\ee
$$
\rightarrow
 (b_8-b_7){\rm log}\mu^2{\rm Im}T_A~{\rm Im}U_A.
$$
Modulo the artificial sub-leading logarithmic contribution (due to
the infrared divergences), the moduli contribution of the second and
third plane $T_A,~U_A,~ A=2,3$, is exponentially suppressed due to
the asymptotic behaviour of $\th_4(T_A),~\th_4(U_A)$ for large $T_A$
and $U_A$,
$ \th_4(T_A~)=1+{\rm \cal O}(e^{-i\pi~T_A })$.

There is a class of such models obtained from $N=2$ $Z_2$
orbifold compactifications
by using $D_4$ type symmetries that act on the twist fields as well
as the lattice.

\section{Type II duals of heterotic groundstates with Spontaneously
Broken Supersymmetry}
\setcounter{equation}{0}

The heterotic string compactified on $T^4$ with N=2 (6-d) spacetime
supersymmetry
has been conjectured to be dual to type II theory compactified on $K_3$
\cite{duff,HT}.
This duality changes the sign of the dilaton, dualizes the
field strength of the antisymmetric tensor and leaves the (4,20) gauge
fields $A^{I}_{\mu}$, the (4,20) moduli
matrix $M_{4,20}$
and the Einstein metric invariant.
Obviously this duality descends to 4-d upon compactifying both theories
on an extra $T^2$.
In four dimensions there are four extra gauge fields, two coming from
the
metric $A^i_{\mu}$ whose charges are the momenta of the $T^2$ and two
coming from
the antisymmetric tensor $B_{i,\mu}$ whose charges are the winding
numbers of the $T^2$.
Moreover we have three extra scalars from the components of the metric
on $T^2$, $G_{ij}$ and one from the antisymmetric tensor $B_{ij}$.
There are also 2$\times$ 24 extra scalars, $Y^i_I$ coming from the 6-d
gauge bosons.
Moreover we can dualize in four dimensions the antisymmetric tensor to
an axion field $A$.
If we denote heterotic variables by unprimed names and type II ones by
primed
names then the heterotic-type II duality in four dimensions implies
that

\be
e^{-\phi}=\sqrt{{\rm det}G'_{ij}}\;\;\;,\;\;\;
e^{-\phi'}=\sqrt{{\rm det}G_{ij}}
\label{d1}\ee
\be
{G_{ij}\over \sqrt{{\rm det}G_{ij}}}={G'_{ij}\over \sqrt{{\rm
det}G'_{ij}}}\;\;\;,\;\;\;A'^{i}_{\mu}=A^{i}_{\mu}
\label{d2}\ee
\be
e^{-\phi}g_{\mu\nu}=e^{-\phi'}g'_{\mu\nu}\;\;\to\;\;g^{E}_{\mu\nu}=
g'^{E}_{\mu\nu}
\label{d3}\ee
\be
M'_{4,20}=M_{4,20}\;\;\;,\;\;\;A^{I}_{\mu}=A'^{I}_{\mu}\;\;\;,\;\;\;
Y^{i}_{I}=Y'^{i}_{I}
\label{d4}\ee

\be
A={1\over 2}\e^{ij}B'_{ij}\;\;\;,\;\;\;
A'={1\over 2}\e^{ij}B_{ij}
\label{d6}\ee
Moreover it effects an electric-magnetic duality transformation on
the $B^i_{\mu}$ gauge fields
\be
{1\over 2}{{\e_{\mu\nu}}^{\rho\sigma}\over
\sqrt{-{\rm det}g}}\e^{ij}F^{B'}_{j,\rho\sigma}
={\delta S^{het}\over \delta F^{B,\mu\nu}_{i}}
\label{d7}\ee
On the electric and magnetic charges it acts as in (\ref{a24a}) on the
$T^2$
charges and leaves the rest invariant.

For the configurations of moduli we are interested in our paper, namely
the factorization $(6,22)\to (2,18)\times (4,4)$, using (\ref{a26}),
(\ref{a27}) we proceed as follows.
In the case of the heterotic string the complex moduli $S,T,U,\vec W$
are defined in terms of the $\sigma$-model data as described in section
2.
However for the type II string the situation is different.
A careful analysis of the tree level action shows that there is an
analogue
of the Green--Schwarz term $B\wedge F\wedge F$ at tree level
\footnote{This appears
at one loop at the heterotic side for both 4-d descendants of $B\wedge
F^4 $ and $B\wedge R^4$. The $B\wedge R\wedge R$ term appears  at one
loop in the type II side \cite{vw2}.}.
This term changes at tree level the definition of the type II $S'$
field.
There is an analogous phenomenon which changes also at tree level the
definition of the $T'$ field.
The correct formulae read:
\be
S'=A'-{1\over 2}Y^I_1Y^I_2+{U_1\over
2}Y^I_2Y^I_2+i(e^{-\phi'}+{U_2\over 2}Y^I_2Y^I_2)
\label{e5}\ee
\be
T'=\sqrt{detG'_{ij}}+iB'
\ee
where as usual
\be
{1\over \sqrt{det~G_{ij}}}G'_{ij}={1\over U_2}\left(\matrix{1&U_1\cr
U_1&|U|^2\cr}\right)
\label{e6}\ee
Thus (\ref{d1})--(\ref{d6}) translate to
\be
U=U'\;\;\;,\;\;\;\vec W=\vec W'
\label{d16}\ee
\be
S=T'\;\;\;,\;\;\;T=S'
\label{d14}\ee

Let us indicate how the N=4 heterotic-type II duality works at the
level of our restricted (2,18) BPS formula given in (\ref{a36}).
Let us start from the heterotic string not necessarily weakly coupled.
We would like however to end up and compare with the weakly coupled
type II string.
Thus we must take the limit $T_2$ large in the mass formula and keep
light states
\be
M^2_{het}={|-m_1+m_2 U+\vec W\cdot \vec Q+S(-\tilde m_{1}U+\tilde
m_{2}+\vec W\cdot\tilde {\vec Q})|^2\over 4~S_2(T_2U_2-(\vec W_2)^2/2)}
\label{e1}\ee
Using type II variables from (\ref{d14}) we can
write (\ref{e1}) as

\be
M^2_{pert-II}={|-m_1+m_2 U'+\vec W'\cdot \vec Q+T'(-\tilde
m_{1}U'+\tilde m_{2}+\vec W'\cdot \tilde{\vec Q})|^2\over
4~\left(S'_2-{\vec {W'}_2^2\over 2U'_2}\right)T'_2U'_2}\label{e2}\ee
This gives the $almost$ correct tree-level type II mass formula in the
large $T'_2$ limit, taking into account (\ref{e5}) and the duality map
(\ref{a24a}).
In type II perturbation theory there are no charged states coupled to
the Wilson lines.
However such states seems to appear in the perturbative formula.
The reason is that although such states are not visible in type II
perturbation
theory their mass is not suppressed in perturbation theory.
This is similar to what is expected to happen in conifold transitions
\cite{con}.

Orbifolding both sides by the same freely acting symmetry we will
obtain
new dual groundstates, due to the adiabatic argument of \cite{vw}.
Thus we would like to identify the duals of the heterotic models
constructed
in the previous sections with spontaneously broken supersymmetry.

For concreteness we will go to the $Z_2$ submanifold of $K_3$ where the
conformal field theory is explicit and we will map directly the
heterotic to
the type II string.
The type II partition function on $K_3\times T^2$ at the orbifold point
is
\be
Z^{II}_{N=4}={1\over
8}\sum_{\a,\b=0}^{1}\sum_{\bar\a,\bar\b=0}^{1}\;\sum_{h,g=0}^{1}
(-1)^{\a+\b+\a\b}{\vartheta^2[^{\a}_{\b}]
\vartheta[^{\a+h}_{\b+g}]\vartheta[^{\a-h}_{\b-g}]\over
\eta^4}\;\;\times
\label{d9}\ee
$$\times (-1)^{\bar\a+\bar\b+\bar\a\bar \b}{\bar
\vartheta^2[^{\bar\a}_{\bar\b}]
\bar\vartheta[^{\bar\a+h}_{\bar\b+g}]
\bar\vartheta[^{\bar\a-h}_{\bar\b-g}]\over \bar\eta^4}\;\;
{1\over {\rm Im}\tau
|\eta|^4}\;\;{\Gamma_{2,2}\over
|\eta|^{4}}Z_{4,4}^{\rm twist}[^h_g]
$$
Let us look at the massless bosonic spectrum and match it to that of
the N=4 heterotic string.
Consider first the NS-NS sector $(\a=\bar\a=0)$.
In the untwisted sector (h=0) there are  32 degrees of freedom
corresponding to the graviton, 2 scalars (axion-dilaton), 4 vectors,
and another 20 scalars
(the $\Gamma_{2,2}$ and $Z^{\rm twist}_{4,4}$ moduli).
Two of the gauge bosons are graviphotons while the other two belong to
U(1) vector multiplets. Thus these four gauge bosons have lattice
signature (2,2).
Similarly the (2,2) moduli belong to these two vector multiplets
while the (4,4) moduli are in multiplets with vectors coming from the
R-R untwisted sector.
In the twisted sector (h=1) there are 16 $Z_2$ invariant ground states
in
the $T^4/Z_2$ part: $H^{I}$.
There are in total $4\times 16$ massless states,
all of them scalars belonging to vector multiplets along with vectors
coming
from the R-R twisted sector.

In the R-R $(\a=\bar\a=1)$ untwisted sector there are 32 physical
degrees of freedom.
These correspond to 8 vectors and 16 scalars.
The vectors have lattice signature (4,4) and four of them are
graviphotons
while the other four are in vector multiplets.
The sixteen scalars complete the six vector multiplets.

Finally in the R-R, twisted sector, there are $4\times 16$ massless
states corresponding to 16 vectors and 32 scalars.

Here the gauge group is composed of U(1)'s which implies that we are
at a generic point in the space of Wilson lines.
The perturbative spectrum is charged under two of
the graviphotons and two of the other gauge bosons with charges given
by $p_L,p_R$ of the $T^2$.

Consider now the freely acting orbifold groundstates on the heterotic
side
that consisted of a rotation on the (4,4) part of the lattice
and a (2,18) translation. Again for simplicity we focus on the $Z_2$
case.
The $Z_2$ rotation on the (4,4) part changes on the type II side the
sign
of the massless states which come from the untwisted R-R sector as well
as the scalars coming from the twisted NS-NS sector.
The effect of the (2,18) translation $\e=(\vec\e_L;\vec\e_R,\vec\zeta)$
is to give phases to massive charged states, but has no effect on the
massless spectrum.
Thus at the massless level the NS-NS twisted and R-R untwisted sectors
have to be projected out.
The projection in the type II  case which has the same effect as the
(4,4) rotation in the heterotic side is a combination of
$(-1)^{F_{R}}$,
which changes the sign of the right-moving Ramond sector, and the
symmetry transformation $e$
described in Appendix A, which acts on the twisted ground states of the
orbifold
with a minus sign and is inert on anything else.

The $\vec\zeta$ translation vector does not act in the perturbative
type II string
since the perturbative spectrum does not contain states charged under
the 16
gauge bosons coming from the R-R twisted sector.
However it will act on non-perturbative D-brane states carrying R-R
charges.
Finally the phase coming from the translation of the (2,2) piece is

\be
(-1)^{\vec m\cdot\e_{R}+\vec n\cdot\e_{L}}
\label{d11}\ee
in the heterotic side.
Under the type II-heterotic map (\ref{a24a}) it becomes in the
heterotic side
\be
(-1)^{\vec m\cdot\vec\e_{R}+\tilde{\vec m}\hat\times\vec \e_{L}}
\label{d12}\ee
where $\vec a\hat\times \vec b=a_1b_2-a_2b_1$.
Thus the $\e_L$ translation acts on the type II side on the
magnetically charged states of the momentum-gauge fields of the
2-torus and thus is also not visible in type II perturbation theory.
The type II duals have 20 vector multiplets and 4 hypermultiplets
and are thus ``mirrors" of the type II grounstate  discussed in
\cite{vw}
with 4 vector multiplets and 20 hypermultiplets.

Thus, at the perturbative type II level the partition function for the
models dual to the heterotic ones is

\be
Z^{II}_{N=4\to N=2}={1\over
16}\sum_{\a,\b=0}^{1}\sum_{\bar\a,\bar\b=0}^{1}\;\sum_{h,g,\bar h\bar
g=0}^{1}
(-1)^{\a+\b+\a\b}{\vartheta^2[^{\a}_{\b}]
\vartheta[^{\a+h}_{\b+g}]\vartheta[^{\a-h}_{\b-g}]\over
\eta^4}\;\;\times
\label{d10}\ee
$$\times (-1)^{\bar\a+\bar\b+\bar\a\bar \b+(\bar\a+h)\bar g+(\bar
b+g)\bar h+\bar g\bar h}{\bar
\vartheta^2[^{\bar\a}_{\bar\b}]
\bar\vartheta[^{\bar\a+h}_{\bar\b+g}]
\bar\vartheta[^{\bar\a-h}_{\bar\b-g}]\over \bar\eta^4}\;\;
{1\over {\rm Im}\tau
|\eta|^4}\;\;{\Gamma^{\vec \e_R}_{2,2}[^{\bar h}_{\bar g}]\over
|\eta|^{4}}Z_{4,4}^{\rm twist}[^h_g]
$$

Here the reader might have noticed a potential puzzle.
Consider a heterotic ground state that contains a translation with
$\vec\e_{R}=\vec 0$.
In such a groundstate, in the limit ${\rm Im}T\to 0$ N=2 supersymmetry
is restored to N=4. Alternatively speaking $m_{3/2}\sim {\rm Im}T$.
Thus, in weakly-coupled heterotic string, we take $S\to \infty$
and also $T\to 0$.
According to our duality map described above, there is no perturbative
shift of the $T^2$ in the type II side.
Thus, in perturbation theory the type-II ground-state does
not look like a spontaneously broken N=4 groundstate.
However a look at (\ref{d12}) is enough to convince us that
there are two  gravitini, with $m_{3/2}\sim {\rm Im}S'$ which are
light in the strong coupling region of the type II theory and
certainly
not visible in the weak coupling type II perturbation theory.

A similar phenomenon can happen in reverse.
Consider a freely acting orbifold of the type II (N=4) side as in
(\ref{d10}),
where the (2,2) lattice translation acts on the windings of the
2-torus with the phase $(-1)^{\vec\e_L\cdot \vec n}$.
This is modular invariant on the type II side.
On the heterotic side the shift of the 2-torus becomes
non-perturbative
via the heterotic-type II map (\ref{a24a}), $(-1)^{\tilde{\vec m}\hat
\times \vec\e_{L}}$.
Thus, in heterotic perturbation theory, we only see the $Z_2$ rotation
of
the (4,4) torus. As it stands the heterotic ground state is not modular
invariant. An extra shift in the gauge lattice is needed (not visible
on
the type II side).
Thus the perturbative heterotic ground state has a $K_3\times T^2$
structure
(at the $Z_2$ orbifold point) and the supersymmetry $N=4\to N=2$ is
explicitly
 broken in perturbation theory.
Turning on all Wilson lines we find that the generic massless spectrum
has 19 vector multiplets (including the dilaton) and 4 hypermultiplets.
Moreover the $SL(2,Z)_S$ is broken to $\Gamma^-(2)_S$ as can be easily
seen by following the fate of $T$-duality of the type II dual.

This brings us to analyse the following issue.
It is widely believed that $K_3\times T^2$ compactifications of the
heterotic string (with N=2 spacetime supersymmetry) have type II duals.
In the cases that have been studied \cite{kv}--\cite{qu} the type II
duals
are CY (symmetric) compactifications. One of the spacetime
supersymmetries
comes from
the left-moving sector while the other comes from the right-moving
sector.
Moreover, it has been argued \cite{al} that the CY manifold must be a
$K_3$ fibration.
Let us consider the question whether the $K_3\times T^2$
compactification with the standard embedding of the gauge group into
$E_8$, described by the above orbifold has a type II dual that is a
$K_3$-fibration.
Such a ground state has generically 19 vector multiplets and 4
hypermultiplets.
At the orbifold point with zero Wilson lines the gauge group is
$E_8\times E_7\times SU(2)\times U(1)^3$.
The $K_3$-fibered Calabi-Yau must have $(h_{21},h_{11})=(19,3)$.
It turns out that such a manifold does not exist \cite{akms}.
Our previous argument strongly suggests that the correct type II dual
of
this heterotic compactification is the asymmetric type II groundstate
described above where both supersymmetries come from one side.
In particular this type II ``compactification" does not have a
geometrical interpretation.
The story becomes more intriguing once we first go to the enhanced
symmetry point and subsequently higgs the SU(2).
Then, there is a series of candidate $K_3$-fibrations in the list of
\cite{akms}
describing a sequence of ground states, obtained from
the original one by sequential Higgsing.\footnote{The final four were
already found in \cite{qu} based on the list given in \cite{klm}.}
We have the following sequence
$$
(h_{21},h_{11})=(18,64)\to(17,83)\to(16,100)\to(15,115)\to(14,166)\to
$$
$$
\to(13,229)\to(12,318)\to(11,491)
$$
These correspond to the cascade breaking \cite{kv,qu}
$$
E_7\to E_6\to SO(10)\to SU(5)\to SU(4)\to SU(3)\to SU(2)\to 0
$$
This strongly suggests that the higgsing of the SU(2) on the heterotic
side corresponds to a non-perturbative transition between the original
asymmetric type II vacuum to a symmetric one described by the (18,64)
$K_3$-fibration.
In reverse, this CY manifold should have a singular limit where an
SU(2) symmetry appears. At this point a new region of moduli space is
opening where
there is no longer  any geometrical interpretation, the ground state
being described by an asymmetric CFT.
In some respects this looks like the conifold transition but its
interpretation
seems to be even more exotic.
It would be very interesting to quantitatively test this picture.

Another comment concerns the fate of the $SL(2,Z)_S$ electric-magnetic
duality symmetry of the original N=4 theory, in the spontaneously
broken phase.
It is known that in the N=4 case $SL(2,Z)_S$ is a corollary of
heterotic-type II duality, since the $T$-duality of type II translates
into the S-duality of the heterotic theory.
Let us investigate what remains of the perturbative $T$ duality in the
broken type II vacuum.
We have argued above that the 2-torus on the type II side gets a
(perturbative) shift
$(\vec 0;\vec\e_R)$ which amounts to the phase $(-1)^{\vec
m\cdot\vec\e_{R}}$.
The $SL(2,Z)_T$ acts on the 2-torus charges as the set of matrices
\be
SL(2,Z)_T\;\;\;:\;\;\;\left(\matrix{\vec m\cr\vec n\cr}\right)\to
\left(\matrix{a~{\bf 1}&b~i\sigma^2\cr -c~i\sigma^2&d~{\bf
1}\cr}\right)
\left(\matrix{\vec m\cr\vec
n\cr}\right)\;\;,\;\;ad-bc=1\;\;,\;\;a,b,c,d\in Z
\label{d13}\ee
There are two subgroups of $SL(2,Z)$ that are relevant in this paper.
One is $\Gamma^+(2)$ defined by $b$ even in (\ref{d13}), the
other one is $\Gamma^-(2)$ defined by $c$ even in (\ref{d13}).
Thus when $\vec\e_R\not=\vec 0$, $SL(2,Z)_{T}$ is broken to
$\Gamma^+(2)_T$.
Thus, the $S$-duality group of these ground states is $\Gamma^+(2)_S$.

In the above discussion, it is obvious that there are non-perturbative
ambiguities in the translation related projections.
The most general projection conceivable is determined by our
``electric"
translation vector $\e$, but simultaneously by a ``magnetic"
translation vector $\tilde \e$ whose effects are not visible in the
perturbative spectrum.
Parts of these translations are never perturbatively visible either on
the
heterotic or on the type II side.
We will comment more on this issue in the next section.

One more remark is in order about the type II duals described above.
Inspection shows that all of the N=2 spacetime supersymmetry comes
from the left side. Consequently, in these models the S field is in a
vector multiplet \cite{vw}.
Thus, like in the heterotic side, the vector-moduli space gets
corrections while the hypermultiplet moduli space does not.
At generic Wilson lines this class of models has a massless spectrum
which consists, apart from the supergravity and the dilaton vector
multiplet,
of eighteen  vector multiplets and four neutral hypermultiplets (the
moduli of the 4-torus).
The non-perturbatively exact hypermultiplet quaternionic manifold is
$SO(4,4)/SO(4)\times SO(4)$.
The exactness of the hypermultiplet moduli space restricts the
orbifolding possibilities on the type II side to the ones described in
(\ref{d10}).

The observations made above suggest  the intriguing possibility  that
all heterotic groundstates with N=2 (or even less) spacetime
supersymmetry in four dimensions always have massive gravitini in the
full non-perturbative spectrum.
All such groundstates would have the features of a spontaneously broken
N=4 theory once non-perturbative corrections are taken into account.

\section{Non-perturbative BPS spectrum in partially broken SUSY $N=4
\to N=2$ }
\setcounter{equation}{0}

Following the philosophy of \cite{dvv} we can make a conjecture for the
non-perturbative multiplicities of BPS states in the groundstates we
discussed with spontaneously broken $N=4\to N=2$ supersymmetry.
This consists in
generalizing  the perturbative multiplicity functions
(\ref{bb47})--(\ref{bb49}) $F_i$ to  genus-2 forms.
First we rewrite $F_i$ in a more convenient form:
\be
F_1 =   {1 \over {\bar \eta}^{24}}~ \chi \left[^0_1\right],~~
F_{\pm}={1 \over {\bar \eta}^{24}}~
\left({\bar \chi} \left[^1_0\right]\pm{\bar \chi}
\left[^1_1\right]\right)
\ee
where ${\bar \chi}\left[^h_g\right]$ are given in terms of the
characters of four twisted 2d right-moving bosons:
\be
{\bar \chi}\left[^h_g\right]~
=~{4(-)^h~{\bar \eta}^6 \over {\bar
\th}[^{1+h}_{1+g}]~{\bar\th}[^{1-h}_{1-g}]},
\ee
where in the above equation $(h,g)\ne 0$. We can extend the validity
of ${\bar \chi}\left[^h_g\right]$  for all  $(h,g)$ sectors using
identities between right-moving,  bosonic and fermionic, ``twisted"
characters:
\be
{\bar \chi}\left[^h_g\right]={1\over 8~{\bar
\eta}^6}~\sum_{a,b}(-)^h~{{\bar\th}^4[^{a+h}_{b+g}]~
{\bar\th}^4[^{a-h}_{b-g}]~
{\bar\th}[^{1+h}_{1+g}]~
{\bar\th}[^{1-h}_{1-g}]}.
\label{char}\ee
In this expression, the absence of the $(h,g)=(0,0)$ sector is due to
the vanishing of the odd-spin structures (${\bar\th}[^1_1]$ terms).
At genus-2
$h$ and $g$ become ${\vec h}=(h,~{\tilde h})$ and ${\vec
g}=(g,~{\tilde g})$ in correspondence with the ``electric" and
``magnetic" charge shifts. The generalization in genus-2 of the
twisted characters consists in promoting
the various $\th$-functions with characteristics to their  genus-2
form
\be
{\bar\th}[^{a~+~h}_{b~+~g}]({\bar \tau})~~~\to~~~ {
\bar\th}\left[^{{\vec
a}~+~{\vec h}}_{{\vec b}~+~{\vec g}}\right]({\bar \tau}^{ij}).
\ee
Then, the proposed non-perturbative multiplicities will be generated
by the genus-2  functions:
\be
F \left[^{\vec h}_{\vec g}\right]~=~{ \Phi({\bar \tau}^{ij})}~{\bar
\chi}\left[^{\vec h}_{\vec g}\right]({\bar\tau}^{ij}),
\ee
where $\Phi({\bar \tau}^{ij})$ is the $N=4$ multiplicity function and
 ${\bar \chi}\left[^{\vec h}_{\vec g}\right]({\bar \tau}^{ij})$ are
the genus-2 analogues of the genus-1 ``twisted" characters
${\bar \chi}[^h_g]({\bar \tau})$ defined above.

Using the genus-2 multiplicity functions, we can construct  weighted
free-energy supertraces, which extend at the  non-perturbative level
the same perturbative quantities, e.g. the moduli dependence of the
gauge and gravitational couplings. We define by ${\cal L}^{\cal D}$
the following quantity:
$$
{\cal L}^{\cal D}=\int_{\cal C} [dt\prod
dX^{ij}]~\sum_{h_i,g_i,q_i}{\cal D}(\tau^{ij})~
F \left[^{\vec h}_{\vec g} \right] ({\bar \tau}^{ij})~{\rm
exp}\left[-2i\pi~{\rm Re}{\tau} ^{ij}~({\vec q}_i+{\vec
\e}_i)\cdot({\vec
q}_j+{\vec \e}_j) ~\right]
$$
\be
\times~{\rm exp}~\left[-\pi~t~M^2_{BPS}(S;~{\vec q}_i,{\vec \e}_i)~
 \right],
\ee
where $M^2_{BPS}(S;~{\vec q}_i,{\vec \e}_i)$ stands for the
non-perturbative
mass formula (\ref{a36}) with shifted charges;
$M^2_{BPS}$  depends on the  shifted ``electric" and ``magnetic"
charges, the moduli $T, U,$ and ${\vec W}$ as well as the
dilaton--axion  modulus field $S$.
 The ``period"  matrix $\tau^{ij}$ in eq. (\ref{tauij}),  is
constructed in terms of  the  parameters $t$, $X^{ij}$ and  $S$ in
the following way:
\be
t=\sqrt{{\rm det}(\tau^{ij})},~~~
X^{ij}={\rm Re}~\tau^{ij},~~~{\rm and}~~~
{ \tau^{ij}\over \sqrt{det~\tau^{ij}} }={1\over {\rm
Im}S}\left(\matrix{1&{\rm Re}S\cr {\rm Re}S&|S|^2\cr}\right).
\label{tauij}\ee
The integration on $X^{ij}$ in the domain $[-1/2,~+1/2]$ would give
rise to the non-perturbative matching conditions (\ref{bb42}).
The  relevant multiplicities are generated by the functions  $F
\left[^{\vec h}_{\vec g} \right]$.
This is a suggestive argument, and stands on a similar footing with
the analogous $\tau_1$ integration in the perturbative string.
However we suggest that, like in the string case,  the correct
integration domain is the genus-2  fundamental region.
Thus we expect that the integration over $t$ (in the fundamental
domain of genus-2 with $S$ fixed) gives rise to the non-perturbative
quantity ${\cal L}^{\cal D}[S;~T,~U,~{\vec W}]$ in terms of all
moduli, $S$ included.

The kernel ${\cal D}$ is the non-perturbative analogue of a product of
charge
operators. In the perturbative string, this is given by a product of
right-moving lattice vectors and contains also a ``back-reaction"
term \cite{kk}.
There is an analogue of ``right-moving" charges in the
non-perturbative case when we also include the magnetic charges.
The charge sum for the overall trace can be written in the
perturbative case as a $\bar\tau$ derivative, which would generalize in
the non-perturbative treatment to the
$\partial_{\tau^{11}}+\partial_{\tau^{22}}$.
The  ``back-reaction'' term  can be fixed since it has to
restore the modular properties of the ${\bar \tau}^{ij}$ derivatives.

The physical interpretation of the summation over the ``magnetic"
charges reproduces the Euclidean space-time instanton corrections
to the couplings.

The determination of the non-perturbative effective coupling
constants (the gravitational one included) defines
the non-perturbative prepotential of  the  $N=2$ effective
theory. Therefore, the knowledge of ${\cal L}^{\cal D}_{l}$
determines   at the non-perturbative level
the $N=2$ low-energy  effective  supergravity,  which includes terms
up to two derivatives.

\section{Outlook}

We have demonstrated the existence of partial spontaneous
supersymmetry
breaking in string theory, and gave several concrete examples in both
 the heterotic and type II theories.
We have studied the issue of restoration of supersymmetry, at the
classical and perturbative level.
We have further analysed the consequences of heterotic--type II
duality
valid for the $N=2$ models we presented. We have pointed out that in
the
dual theories the $N=4\to N=2$ supersymmetries may look  explicitly
broken in their perturbation theory.
This was also corroborated by our conjecture on the full
non-perturbative structure of their effective theories.
In some cases we can predict some novel non-perturbative
(non-geometric) transitions
between vacua of the type II string with (2,0) and (1,1) space-time
supersymmetry.

An analysis of the perturbative BPS states of strings,
with supersymmetry spontaneously broken $N=4\to N=2$, and the
underlying
duality structure permit us to conjecture the full non-perturbative
form of the effective field theory.
This conjecture needs to be elaborated and tested in the context of
explicit models. This will be the subject of future analysis.

\vskip 2cm

\centerline{\bf Acknowledgements}

We would like to thank, M. Kreuzer, F. Quevedo and E. Verlinde
for discussions.
C. Kounnas was  supported in part by EEC contracts
SC1$^*$-0394C, SC1$^*$-CT92-0789 and ERBFMRX-CT96-0090.
E. Kiritsis was supported in part by the EEC contract
ERBFMRX-CT96-0090.

\vskip 2cm

\renewcommand{\theequation}{A.\arabic{equation}}
\centerline{\bf\large Appendix A: Orbifold Blocks}
\setcounter{equation}{0}

In this appendix we will derive various $Z^2$ orbifold blocks relevant
for the partition functions of ground states with spontaneously broken
supersymmetry.
Consider the 2-torus lattice sum
\be
\Gamma_{2,2}(T,U)=\sum_{(\vec m;\vec n)\in (Z^2;Z^2)}q^{{1\over
4}p_L^2}
\bar q^{{1\over 4}p_R^2}
\label{A1}\ee
Let us first consider the blocks of the orbifold generated by a
non-trivial $Z_2$ translation given by one-half the  lattice vector
$\e=(\vec\e_L,\vec\e_R)$, whose components are composed of zeros and
ones.
Under such a translation, the U(1) current oscillators are invariant
while
the ground states transform as:
\be
|\vec m;\vec n\rangle\to e^{\pi i (\vec\e_L\cdot\vec
n+\vec\e_R\cdot\vec m)}
|\vec m;\vec n\rangle
\label{A2}\ee
Then the projected partition function is
\be
Z^{\e}_{2,2}[^1_t]={1\over |\eta|^4}\sum_{(\vec m;\vec n)\in (Z^2;Z^2)}
e^{\pi i (\vec\e_L\cdot\vec n+\vec\e_R\cdot\vec m)}q^{{1\over 4}p_L^2}
\bar q^{{1\over 4}p_R^2}
\label{A3}\ee
where $t$ stands for the translation element satisfying $t^2=1$.
In the twisted sector states are in one-to-one correspondence with
lattice
states
with $(\vec m;\vec n)$ shifted by $\e/2$.
Thus,

\be
Z^{\e}_{2,2}[^t_1]={1\over |\eta|^4}\sum_{(\vec m;\vec n)\in
(Z^2;Z^2)+\e/2}
q^{{1\over 4}p_L^2} \bar q^{{1\over 4}p_R^2}
\label{A4}\ee
\be
Z^{\e}_{2,2}[^t_t]={1\over |\eta|^4}\sum_{(\vec m;\vec n)\in
(Z^2;Z^2)+\e/2}
e^{\pi i (\vec\e_L\cdot\vec n+\vec\e_R\cdot\vec m)}q^{{1\over 4}p_L^2}
\bar q^{{1\over 4}p_R^2}
\label{A5}\ee
We can summarize the above as
\be
Z^{\e,\rm shift}_{2,2}[^h_g]={1\over |\eta|^4}\sum_{(\vec m;\vec n)\in
(Z^2;Z^2)+h\e/2}
e^{\pi i g(\vec\e_L\cdot\vec n+\vec\e_R\cdot\vec m)}q^{{1\over 4}p_L^2}
\bar q^{{1\over 4}p_R^2}
\label{A17}\ee

Similarly, denote by $r$ a $Z_2$ rotation ($r^2=1$) that acts in the
standard way:
\be
r~:~ a^i_m\to -a^i_m\;\;\;,\;\;\;\bar a^i_m\to -\bar a^i_m\;\;\;,\;\;\;
|\vec m;\vec n\rangle\to |-\vec m;-\vec n\rangle
\label{A6}\ee
We have as usual
\be
Z^{\rm twist}_{2,2}[^h_g]=\Gamma_{2,2}[^{r^h}_{r^g}]=4{|\eta|^2\over
|\th[^{1+h}_{1+g}]|^2}
\label{A7}\ee

There are four ground states in the $r$-twisted sector, the twist
fields
$H^{ij}$,  $i,j=0,1$ which are in one-to-one correspondence with the
fixed
points of the $r$ action on the 2-torus located at the half-period
points.
The twist fields are invariant under the rotation $r$ but do transform
under the translation $t_{\e}$.
To determine the transformation properties let us write $\vec
\e_L=(\e^1_L,\e^2_L)$ and $\vec \e_R=(\e^1_R,\e^2_R)$
and define the matrices
\be
T_{\e}^{1}=\left(\matrix{(-1)^{\e_L^1}(1-\e^1_R)&(-1)^{\e_L^1}\e^1_R\cr
\e^1_R&
1-\e^1_R\cr}\right)\;\;,\;\;T_{\e}^{2}=
\left(\matrix{(-1)^{\e_L^2}(1-\e^2_R)&(-1)^{\e_L^2}\e^2_R\cr
\e^2_R& 1-\e^2_R\cr}\right)
\label{A11}\ee
Then
\be
t_{\e}\;\;:\;\;H^{ij}\to \sum_{k,l}~T^{1}_{ik}~T^{2}_{jl}~H^{kl}
\label{A12}\ee
There is another $Z_2$ transformation which we will denote by $e$ which
acts trivially on all torus states but has a non-trivial action on the
twist fields:
\be
e\;\;:\;\;H^{ij}\to -H^{ij}
\label{A13}\ee

Consider also the element $rt$ which is a product of a rotation and
translation, ($(rt)^2=1$).
In terms of its action on the spectrum it is essentially the same as
the element $r$. The 4 twist fields are now located at the half-periods
shifted
by $\vec\e_R/4$:
\be
\Gamma_{2,2}[^{(rt)^h}_{(rt)^g}]=\Gamma_{2,2}[^{r^h}_{r^g}]
\label{A8}\ee

Let us now consider the remaining blocks.
We have
\be
Z^{\e}_{2,2}[^t_r]=0
\label{A9}\ee
since the $r$ projection gets contributions only from the states with
$\vec m=\vec n=\vec 0$ while the $t$ twist produces states with $(\vec
m;\vec n)\not=(\vec 0;\vec 0)$ as can be seen from  (\ref{A4}).
We also have
\be
Z^{\e}_{2,2}[^r_t]=0
\label{A14}\ee
consistent with modular invariance and (\ref{A9}), due to the
transformation properties
of the twisted ground states (\ref{A12})
\be
Z^{\e}_{2,2}[^{rt}_r]=0
\label{A177}\ee
since the twisted ground states of $rt$ are isomorphic to those of $r$
but localized at quarter points on the torus. The rotation acts by
interchanging them giving zero trace.
Also
\be
Z^{\e}_{2,2}[^{r}_{rt}]=0
\label{A15}\ee
since the translation projection acts non-trivially on the twist
fields.

Finally
\be
Z^{\e}_{2,2}[^{rt}_t]=Z^{\e}_{2,2}[^t_{rt}]=0
\label{A10}\ee

We can summarize the above in the following way:

\be
Z_{2,2}^{\e}[^{h;H}_{g;G}]\equiv Z_{2,2}^{\e}[^{t^hr^H}_{t^gr^{G}}]
\label{A16}\ee
with
\be
Z_{2,2}^{\e}[^{h;0}_{g;0}]=Z^{\e,\rm shift}_{2,2}[^h_g]
\label{A18}\ee
\be
Z_{2,2}^{\e}[^{0;h}_{0;g}]=Z_{2,2}^{\e}[^{h;h}_{g;g}]=Z^{\rm
twist}_{2,2}[^h_g]
\label{A19}\ee
\be
Z_{2,2}^{\e}[^{h;H}_{g;G}]=0\;\;\;,\;\;\;{\rm otherwise}
\label{A20}\ee

In order to construct N=1 models in sections 5,6 the modular properties
of
the blocks above are important.
We will assume that $\vec\e_R=\vec 0$ since this is the case of
interest.
Then
\be
\tau\to\tau+1\;\;:\;\;Z^{\e}_{2,2}[^{h;H}_{g;G}]\to
Z^{\e}_{2,2}[^{h;H}_{g+h;G+H}]
\label{A21}\ee
\be
\tau\to-{1\over \tau}\;\;:\;\;Z^{\e}_{2,2}[^{h;H}_{g;G}]\to
Z^{\e}_{2,2}[^{g;G}_{h;H}]
\label{A22}\ee

\end{document}